\documentclass[a4paper, 12pt]{scrartcl}

\usepackage{amsmath}
\usepackage{amsfonts}
\usepackage{amssymb}
\usepackage{amsthm}
\usepackage{bbold}
\usepackage{bm}
\usepackage{caption}
\usepackage{url}
\usepackage{float}
\usepackage{tikz}
\usetikzlibrary{decorations.pathreplacing}

\usepackage{accents}
\newcommand*{\m}{\bm{m}}
\newcommand*{\ud}{\bm{u}}
\newcommand*{\pd}{\bm{p}}
\newcommand*{\dtot}{\text{d}}
\newcommand*{\heff}{\bm{H}}
\newcommand{\bnabla}{\bm{\nabla}}
\newcommand{\intt}{\int_{t_0}^{t_1}}

\begin{document}
\title{Modeling Magnetoelastic Wave Interactions in Magnetic Films and Heterostructures: A finite-difference approach}

\author{Peter Flauger$^{1,2}$, Matthias Küß$^{3}$, Michael Karl Steinbauer$^{1,2,4}$,\\ Florian Bruckner$^1$, Bernhard Emhofer$^1$, Emeline Nysten$^5$, \\Matthias Weiß$^5$, Dieter Suess$^{1,2}$, Hubert J. Krenner$^5$, \\Manfred Albrecht$^3$, Claas Abert$^{1,2}$}

\maketitle

\textsuperscript{1} Faculty of Physics, University of Vienna, Physics of Functional Materials, 1090 Vienna, Austria

\textsuperscript{2} University of Vienna, Research Platform MMM Mathematics-Magnetism-Materials, 1090 Vienna, Austria

\textsuperscript{3} Institute of Physics, University of Augsburg, 86135 Augsburg, Germany 

\textsuperscript{4} University of Vienna, Vienna Doctoral School in Physics, 1090 Vienna, Austria

\textsuperscript{5} Physikalisches Institut, Universität Münster, 48149 Münster, Germany


\section*{Abstract}
The (inverse) magnetostrictive effect in ferromagnets couples the magnetic properties to the mechanical stress, allowing for an interaction between the magnetic and mechanical degrees of freedom. In this work, we present a time-integration scheme for the self-consistent simulation of coupled magnetoelastic dynamics within the framework of finite-difference micromagnetism. The proposed implementation extends the Landau-Lifshitz-Gilbert equation by a strain-induced effective field and concurrently solves the elastic equation of motion, while correctly incorporating stress and strain discontinuities at material interfaces. We then present a comprehensive set of examples, ranging from static stress configurations over material boundaries to simulations of surface acoustic wave attenuation in magnetically structured thin and thick films. These computational experiments both validate the implementation and underscore the importance of properly handling jump and boundary conditions in magnon-phonon interaction studies.

\section{Introduction}
In all ferromagnetic materials, the magnetic properties depend on the mechanical strain and stress. This is known as the magnetoelastic effect. Conversely, the state of the magnetization in a ferromagnet is associated with a magnetic strain. This effect is called magnetostriction. In some ferromagnetic materials, this bi-directional coupling is strong enough to be functionalized, with immediate technical applications found e.g. in electronic article surveillance \cite{Herzer_2003_JMMM}. The effect can also be used as a route to further couple the magnetization to other properties. By attaching a piezoelectric component to a strongly magnetoelastic material, the magnetization can be influenced by an electric signal \cite{Zheng_2004_Science, Goennenwein_2008_PhysStatusSolidi, Bhiler_2008_PhysRevB, Weiler_2009_NewJPhys}. Such a composite multiferroic can be operated at room temperature, overcoming a key limitation of magnetoelectric materials \cite{Eerenstein_2006_Nature}. In the magnetic sensing application of reference~\cite{Kittmann_2018_SciRep}, a set of interdigital transducers (IDTs)  is used to excite shear-horizontal surface acoustic waves (SH-SAW) in an piezoelectric substrate, and this acoustic wave then couples to spin waves (SWs) in an ferromagnetic film sputtered onto the substrate. The underlying mechanics of this magnonic-phononic interaction, that result in angular dependent transmission losses of the surface acoustic waves (SAWs) at resonant fields, have been study in great detail for magnetic thin films~\cite{Dreher_2012_PhysRevB, Kuess_2020_PhysRevLett, Kuess_2021_PhysRevAppl, Hwang_2024_PhysRevLett}. In these studies, the magnetization dynamics is assumed to be homogeneous in the out-of-plane direction and the transmission losses are estimated from the Gilbert losses of the magnetic system. While the general form of SAWs might be known or can be obtained from an eigenmode solver, the back-action of the magnetization on them is usually modeled only be reducing the amplitude in accordance with the lost energy. Indeed, the magnetization dynamics for such a system obtained from the Landau-Lifshitz-Gilbert (LLG) equation including an explicitly time dependent magnetoelastic field term will end up in a pseudo-static state, in which the energy gains from the driving field are compensated by the Gilbert losses \cite{Steinbauer_2025_arXiv}.\\
In this work, we present a computational scheme for the self-consistently coupled dynamics of the magnetization and the displacement. This solver is conceived as an extension of the pytorch \cite{Paszke_2019_arXiv} based finite-differences micromagnetic simulation library magnum.np \cite{Bruckner_2023_SciRep}, and is included in magnum.np's core since version 2.1.0 \cite{magnumnp_gitlab}.\\
The current interest in the magnetoelastic interaction within the field has already spawned multiple codes that do not rely on precomputed or measured mechanical stresses for the magnetoelastic field \cite{Bur_2011_JApplPhys}, but are capable of concurrently computing the magnetization dynamics and the elastodynamics. Proposed self-consistent time integration routines have varying degrees of complexity, ranging from specialized solutions like for 1D magnetic ribbons in finite differences presented in~\cite{Bergmair_2014_JApplPhys} to general routines using the finite elements method in the time domain~\cite{Liang_2014_Nanotechnology, ChenC_2017_ApplPhysLett} or frequency domain \cite{Li_2017_JApplPhys}. Indeed, the introduction of mechanical properties to the scope of the problems makes the finite element method a natural first choices. However, due to the expected large problem sizes, the simple, rectangular geometries and the heavy use of periodic boundary conditions, we will treat the problem in finite differences instead. A notable previous attempt within finite differences was undertaken by mumax+ \cite{Moreels_2025_arXiv, Vanderveken_2021_OpenResEur}. A major challenge for coupled magnetoelastic solvers lies in ensuring the stability of the time-integration scheme and the conservation of energy. In the framework of finite differences in micromagnetism, where the degrees of freedom are usually located in the cell centers and material parameters are assumed to be constant within a cell \cite{Vanderveken_2021_OpenResEur}, this issue is tightly related to open boundary conditions at free surfaces and to jumps in the elastic properties at material interfaces.\\
The implementation of these jump conditions for stress and strain at material interfaces is the main focus of this work. The presented results are meant to illustrate and discuss different aspects of this self-consistently coupled time integration scheme, and move from the computation of static 1D magnetoelastic stress over the excitation of SWs by bulk acoustic shear waves to 2D examples of SAW attenuation in magnetic thin film structures, and layered systems in which the magentic film thickness is in the order of magnitude of the SAW wavelength.    

\section{Theory and Methods}
\subsection{Coupled equations of motion}
We will first derive the coupled set of equations of motion for ${(\m,\ud,\pd)}$. The unit vector field $\m$ is the direction of the magnetization, such that the local magnetization is $\bm{M}=M_s\m$ with $M_s$ being the saturation magnetization. $\ud$ is the displacement and $\pd=\rho\dot{\ud}$ is the momentum density, with $\rho$ denoting the mass density. The dynamics of $\m$ are governed by the LLG

\begin{equation}
\frac{\partial\m}{\partial t} = -\frac{\gamma}{1+\alpha^2}\left[\m\times\heff + \alpha\m\times\left(\m\times\heff\right) \right]
\label{eq:LLG}
\end{equation}

stated here in its explicit form. $\gamma$ is the reduced gyromagnetic ratio, $\alpha$ is the phenomenological Gilbert damping parameter and $\heff$ is the effective field acting on the magnetization. The rate of change $\dot{\m}$ obtained from~(\ref{eq:LLG}) is always perpendicular to $\m$ and components of $\heff$ parallel to $\m$ do not influence the dynamics of $\m$. The effective field is obtained from the functional derivative of the total free energy $E$ with respect to $\m$

\begin{equation}
\heff = -\frac{1}{M_s\mu_0}\frac{\delta E}{\delta\m}.
\label{eq:effective_field}
\end{equation}

The modular total free energy

\begin{equation}
E(\m,\ud,\pd) = T^\text{el}(\pd) + U^\text{el}(\m,\ud) + E^\text{ex}(\m) + E^\text{dem}(\m) + E^\text{aniso}(\m) + E^\text{zee}(\m) + \text{...}
\label{eq:free_energy}
\end{equation}

features the elastic kinetic energy

\begin{equation}
T^\text{el} = \int_{\Omega_u} \frac{\pd^2}{2\rho}\dtot\bm{x}
\end{equation}

and the elastic potential 

\begin{equation}
U^\text{el} = \frac{1}{2}\int_{\Omega_m\cup\Omega_u} \left(\varepsilon-\varepsilon^m\right):C:\left(\varepsilon-\varepsilon^m\right)\dtot\bm{x}
\label{eq:elastic_potential}
\end{equation}

in addition to the usual micromagnetic energy contributions. The elastic potential energy $U^\text{el}$ has an explicit dependence on both $\ud$ and $\m$ and is thus identified as the term that relates the magnetization dynamics to the displacement.   It is stated here in terms of the Frobenius inner product of the elastic stress and elastic strain. The latter is defined as 

\begin{equation}
\varepsilon^\text{el} = \varepsilon-\varepsilon^m
\end{equation}

and is obtained from the total strain 

\begin{equation}
\varepsilon = \frac{1}{2}\left[\bnabla\cdot\ud+\left(\bnabla\cdot\ud\right)^\text{T}\right]
\end{equation}

and the magnetically induced strain, which reads as

\begin{equation}
\varepsilon^m_{ij} = \left\lbrace\begin{array}{ll}
\frac{3}{2}\lambda_{100}\left(m^2_i - \frac{1}{3} \right) & \text{if }i=j \\
\frac{3}{2}\lambda_{111}m_i m_j  & \text{if }i\neq j \\
\end{array}\right.
\end{equation}

for a material with cubic symmetry. The magnetostrictive coupling constants along the $\langle 100\rangle$ and $\langle 111\rangle$ directions being $\lambda_{100}$ and $\lambda_{111}$, respectively. The stress is related to the strain by the $3\times 3\times 3\times 3$ stiffness tensor $C$ via the Frobenius inner product. The total strain is defined as 

\begin{equation}
\sigma = C:\varepsilon
\end{equation}

and the magnetic stress as 

\begin{equation}
\sigma^m = C:\varepsilon^m.
\end{equation}

Since all strains and stresses as well as the stiffness tensor are symmetric, $C$ is composed of only 21 independent elements.\\

Just as the free energy (\ref{eq:free_energy}) can be separated into different contributions, the effective field can be decomposed into field terms corresponding to these energy contributions. This immediately yields a convenient compartmentalization for a micromagnetic simulation package, as it allows to add or remove field terms from the simulation setup in a straight forward manner. We will only quickly go over energy contributions that yield effective field contributions (i.e. that depend on $\m$) and that are relevant to this work. For more details on them, and a derivation of equation (\ref{eq:LLG}) from the Lagrange formalism, we advice the reader to refer to reference~\cite{Abert_2019_EurPhysJB}.\\

The exchange energy  

\begin{equation}
E^\text{ex} = \int_{\Omega_m} A\left(\bnabla\m\right)^2\dtot\bm{x} 
\label{eq:E_ex}
\end{equation}

with exchange stiffness $A>0$ favors a parallel alignment of the magnetization. On the other hand, the demagnetization energy would result in a globally demagnetized magnetic state if no other energy contribution would be present. The demagnetization energy 

\begin{equation}
E^\text{dem} = -\frac{\mu_0}{2}\int_{\Omega_m}M_s\m\cdot\bm{H}^\text{dem}\dtot\bm{x} 
\label{eq:demag_energy}
\end{equation}

is computed using the conservative demagnetization field

\begin{align}
\bm{H}^\text{dem} &= -\bnabla\Phi \\  
&= -\frac{1}{4\pi} \int_{\Omega_m} \bnabla\left(M_s\m(\bm{x}')\cdot\bnabla' \frac{1}{|\bm{x}-\bm{x}'|}\right)\dtot\bm{x}' 
\label{eq:demag_field}
\end{align}

that arises from the magnetic potential $\Phi$ due to the dipole-dipole interaction. The lowest order terms of the crystalline anisotropy energy for uniaxial materials is

\begin{equation}
E^\text{aniso} = -\int_{\Omega_m}K_u\left(\m\cdot\bm{e}_u\right)^2\dtot\bm{x}
\end{equation}

with the uniaxial anisotropy constant $K_u$. When this constant is negative, the anisotropy energy term favors the magnetization to be perpendicular to $\bm{e}_u$ in an easy-plane. 
%
%
The Zeeman energy due to an external field $\bm{H}^\text{ext}$ is given by

\begin{equation}
E^\text{zee} = -\mu_0\int_{\Omega_m}\bm{M}\cdot\bm{H}^\text{ext}\,\dtot\bm{x}.
\end{equation}

The field corresponding to the elastic potential energy (\ref{eq:elastic_potential}) is \cite{Liang_2014_Nanotechnology}

\begin{equation}
H^\text{magEl}_i = -\frac{1}{M_s\mu_0}\left(\sigma-\sigma^m\right) : \frac{\partial\varepsilon^m}{\partial m_i}.
\label{eq:Hmagel}
\end{equation}

Not all terms in (\ref{eq:elastic_potential}) depend on both $\m$ and $\ud$. Instead, after expanding $U^\text{el}$, there is a purely mechanical and a purely magnetic set of terms to be found. We want to point out that in literature, a slightly different linear version of this term is used frequently \cite{ChenC_2017_ApplPhysLett, Yamamoto_2020_JPhysSocJpn, Vanderveken_2021_OpenResEur}. This linear term yields the same torque only for isotropic materials and is not connected to the solely magnetization-dependent terms in the magnetoelastic potential energy~(\ref{eq:elastic_potential}).\\\\
The energy contribution that is related to the field term is the same that lets the magnetic state alter the elastodynamics, and we can see this by deriving the elastic equation of motion from the functional derivatives of the action integral. From the kinetic term, we obtain

\begin{align}
\left.\frac{\dtot}{\dtot\xi}\intt T^\text{el}\left(\dot{\ud}+\xi\dot{\bm{\eta}}\right)\dtot t\right\vert_{\xi=0} &= \left.\frac{\dtot}{\dtot\xi}\intt\int_{\Omega_u}\frac{\rho}{2}\left(\dot{\ud}+\xi\dot{\bm{\eta}}\right)^2\dtot\bm{x}\dtot t\right\vert_{\xi=0}\\
&= \intt\int_{\Omega_u}\rho\dot{\ud}\cdot\dot{\bm{\eta}}\,\dtot\bm{x}\dtot t \\
&= -\intt\int_{\Omega_u}\rho\ddot{\ud}\cdot\bm{\eta}\,\dtot\bm{x}\dtot t
\end{align}

and from the elastic potential energy, 

\begin{align}
\left.\intt\frac{\dtot}{\dtot\xi}U^{\ud}\left(\ud+\xi\bm{\eta}\right)\right\vert_{\xi=0}\dtot t =& \intt\int_{\Omega_u} \left(\sigma-\sigma^m\right):\text{sym}\left(\bnabla\bm{\eta}\right) \dtot\bm{x}\dtot t \\
=& \intt\int_{\Omega_u} \left(\sigma-\sigma^m\right):\bnabla\bm{\eta}\,\dtot\bm{x} \dtot t \\
=& \intt\left[\int_{\partial\Omega_u}\left(\sigma-\sigma^m\right)\cdot\hat{n}\cdot\bm{\eta}\,\dtot \bm{S} - \int_{\Omega_u} \bnabla\cdot \left(\sigma-\sigma^m\right)\cdot\bm{\eta}\,\dtot\bm{x}\right]\dtot t.
\end{align}

To understand how $\text{sym}\left(\bnabla\bm{\eta}\right)$ was reduced to just $\bnabla\bm{\eta}$, one needs to keep the symmetry of the stress tensors in mind. Under the time integral, we have now arrived at the weak form of the elastic problem involving magnetic stresses \cite{Liang_2014_Nanotechnology}.

\begin{equation}
\int_{\Omega_u}\rho\ddot{\ud}\cdot\bm{\eta}\,\dtot\bm{x} = \int_{\Omega_u} \bnabla\cdot \left(\sigma-\sigma^m\right)\cdot\bm{\eta}\,\dtot\bm{x} - \int_{\partial\Omega_u}\left(\sigma-\sigma^m\right)\cdot\hat{n}\cdot\bm{\eta}\,\dtot \bm{S}
\label{eq:LE_weak_from}
\end{equation}

Thus, in a material that occupies the space $\Omega_u$, the elastodynamics can be obtained from the following set of equations \cite{Shu_2004_MechMater}

\begin{equation}
\left\lbrace \begin{array}{ll}
\bm{f}=\rho\ddot{\ud} = \nabla\cdot\left(\sigma-\sigma^m\right) & \text{ on }\Omega_u \\
(\sigma-\sigma^m)\cdot\hat{n} = \bm{t} &  \text{ on }\partial\Omega^t_u \\
\ud = \ud_0 &  \text{ on }\partial\Omega^0_u \\
\end{array} \right.
\label{eq:elastodynamics_equations}
\end{equation}

where $\bm{f}$ is the continuous force density field and $\bm{t}$ is the traction vector, that can be set as a Neumann type boundary condition to match an external pressure. The natural boundary condition of this problem correspond to homogeneous Neumann boundary conditions.\\
In conjunction with the LLG (\ref{eq:LLG}), we obtain a coupled system of equations.

\begin{equation}
\left(\begin{array}{c}
\dot{\m} \\
\dot{\ud} \\
\dot{\bm{p}}
\end{array}\right) = \left(\begin{array}{c}
-\frac{\gamma}{1+\alpha^2}\left[\m\times\heff + \alpha\m\times\left(\m\times\heff\right) \right] \\
\bm{p}/\rho \\
\nabla\cdot\left(\sigma-\sigma^m\right) - \eta\bm{p}
\end{array}\right)
\label{eq:coupled_equations}
\end{equation}

Together with the boundary conditions from (\ref{eq:elastodynamics_equations}), i.e. $\ud=\ud_0$ on $\partial\Omega_u^0$ and $(\sigma-\sigma^m)\cdot\hat{n}=\bm{t}$ on $\partial\Omega_u^t$ as well as the boundary condition $A\bnabla\m\cdot\hat{n}=0$ on $\partial\Omega_m$ obtained for interfaces to non-magnetic materials from the variation of the exchange energy (\ref{eq:E_ex}), this allows us to compute the self-consistent dynamics of $(\m,\ud,\pd)$. In this coupled set of equations (\ref{eq:coupled_equations}), a  phenomenological damping constant $\eta$ \cite{Vanderveken_2021_OpenResEur} was added to (\ref{eq:elastodynamics_equations}), that can be used together with the Gilbert damping in simulation setups to relax the system prior to time-evolution experiments. For the time-integration of (\ref{eq:coupled_equations}), we use the RKF45 solver with adaptive time steps implemented in magnum.np \cite{Bruckner_2023_SciRep, Mathews_2004_Book}.\\
From (\ref{eq:elastodynamics_equations}), we can infer how the magnetization acts on the displacement. The first equation of problem (\ref{eq:elastodynamics_equations}) tells us that bulk forces arise from an inhomogeneous magnetic strain. However, also a homogeneous magnetization can lead to a deformation of the material, since the mechanical stress $\sigma$ and magnetic stress $\sigma^m$ are are coupled due to the Neumann boundary conditions on open surfaces and material interfaces, as we will discuss next.

\subsection{Jump Conditions at Material Interfaces}
\label{subsec:jump_conditions_at_material_interfaces}

When applying the finite differences method to micromagnetism, the magnetization is typically sampled at the center position $(x_i, y_j, z_k)$ of the cells $(\Delta x_i, \Delta y_j, \Delta z_k)$. The degrees of freedom are assumed to be samples of continuous functions and vector fields and the material parameters are assumed to be constant within a cell. These assumptions inform how the approximations of the spatial derivatives of $\ud$ and $\m$ are calculated. For $\m$, the boundary terms obtained from the variation of the exchange energy (\ref{eq:E_ex}) require that $A\bnabla\m\cdot\hat{n}$ is continuous over material interfaces, yielding homogeneous Neumann boundary conditions at interfaces to non-magnetic materials. Between ferromagnetic materials with different exchange constant, this and the in micromagnetism assumed continuous nature of $\m$ can be used to build up a three-point stencil for the spatial derivatives of $\m$ that accounts for these jump conditions, as discussed in reference~\cite{Heistracher_2022_JMMM}. We will now extend this ansatz to model the jump conditions for the mechanical strain components.\\
If the problem region $\Omega_u$ is made up from multiple different materials, then the set of equations (\ref{eq:coupled_equations}) applies to every material subset of $\Omega_u$ and at the interfaces where these materials touch, the traction vector

\begin{equation}
\bm{t}=\left(\sigma-\sigma^m\right)\cdot\hat{n}
\end{equation}

needs to be matched. This can be understood by looking a the weak form (\ref{eq:LE_weak_from}) and introducing boundary terms at the material interfaces that need to compensate each other. We will limit our investigation to materials that can be described by stiffness matrices of the form 

\begin{equation}
C = \left(\begin{array}{cccccc}
C_{11} & C_{12} & C_{13} & 0 & 0 & 0 \\
C_{12} & C_{22} & C_{23} & 0 & 0 & 0 \\
C_{13} & C_{23} & C_{33} & 0 & 0 & 0 \\
0 & 0  & 0 & C_{44} & 0 & 0 \\
0 & 0  & 0 & 0 & C_{55} & 0 \\
0 & 0  & 0 & 0 & 0 & C_{66} 
\end{array}\right)
\label{eq:stiffness_matrix_covered}
\end{equation}

in Voigt notation. This subclass of stiffness matrices is applicable for isotropic, cubic, the (4/mmm) sub-class of tetragonal, hexagonal and orthorombic materials. The individual coefficients need to satisfy the elastic stability conditions \cite{Mouhat_2014_PhysRevB}. Let's consider the case of a material interface perpendicular to $\hat{x}$ at position $x=a$. The individual components of $\bm{t}$ for $\hat{n}=\hat{x}$ are

\begin{align}
t_x &= C_{11} \frac{\partial u_x}{\partial x} + C_{12}\frac{\partial u_y}{\partial y}+C_{13}\frac{\partial u_z}{\partial z} - \sigma^m_{xx} \\
t_y &= C_{66}\left(\frac{\partial u_y}{\partial x}+\frac{\partial u_x}{\partial y}\right) - \sigma^m_{xy} \\
t_z &= C_{55}\left(\frac{\partial u_z}{\partial x}+\frac{\partial u_x}{\partial z}\right) - \sigma^m_{xz}
\end{align}

and from the continuity of $\bm{t}$

\begin{equation}
\lim_{x\rightarrow a}\bm{t} = \lim_{a\leftarrow x}\bm{t}
\end{equation}

we obtain conditions of the form

\begin{equation}
C\frac{\partial u}{\partial x}+B\Bigg\vert_{x=x^-} = C\frac{\partial u}{\partial x}+B\Bigg\vert_{x=x^+}
\label{eq:jump_condition_general_form}
\end{equation}

that become 

\begin{equation}
C_i u'^\rightarrow_i + B^+_i = C_{i+1}u'^\leftarrow_{i+1} + B^-_{i+1}
\label{eq:jump_condition_general_form_FD}
\end{equation}

on an irregular finite differences mesh. Since $\ud$ is continuous, we can require the forward difference $u'^\rightarrow_i$ and the backward difference $u'^\leftarrow_{i+1}$ to lead to the same value of $u$ at the interface at $x=a$.

\begin{align}
u(a) &= u_i + \frac{\Delta x_i}{2} u'^\rightarrow_i \label{eq:u_continuity_l}\\
u(a) &= u_{i+1} - \frac{\Delta x_{i+1}}{2} u'^\leftarrow_{i+1} \label{eq:u_continuity_r}
\end{align}

Note that $a$ is only the half-way point if $\Delta x_i=\Delta x_{i+1}$, i.e if the cells sharing the interface have the same width. Rewriting (\ref{eq:jump_condition_general_form_FD}) gives us

\begin{equation}
u'^\leftarrow_{i+1} = \frac{1}{C_{i+1}}\left(C_i u'^\rightarrow_i + B^+_i-B^-_{i+1} \right)
\label{eq:jump_condition_bwd_diff}
\end{equation}

and after equating with (\ref{eq:u_continuity_l}) and (\ref{eq:u_continuity_r}), we get an expression for the forward difference.

\begin{align}
u_i + \frac{\Delta x_i}{2} u'^\rightarrow_i &= u_{i+1} - \frac{\Delta x_{i+1}}{2 C_{i+1}}\left(C_i u'^\rightarrow_i + B^+_i-B^-_{i+1}\right)  \\
\frac{1}{2}\left(\Delta x_i+\frac{C_i}{C_{i+1}}\Delta x_{i+1}\right)u'^\rightarrow_i &= u_{i+1}-u_i-\frac{\Delta x_{i+1}}{2 C_{i+1}}\left(B^+_i-B^-_{i+1}\right) 
\end{align}

\begin{equation}
u'^\rightarrow_i =  \frac{2C_{i+1}\left(u_{i+1}-u_i\right) + \Delta x_{i+1}\left(B^-_{i+1} - B^+_i\right)}{\Delta x_{i+1}C_i+\Delta x_iC_{i+1}}
\label{eq:fwd_diff}
\end{equation}

inserting (\ref{eq:fwd_diff}) into (\ref{eq:jump_condition_bwd_diff}) and shifting the index back by one yields the backward difference.

\begin{align}
u'^\leftarrow_i =  \frac{2C_{i-1}\left(u_i-u_{i-1}\right) + \Delta x_{i-1}\left(B^+_{i-1}- B^-_i\right)}{\Delta x_{i-1}C_i+\Delta x_iC_{i-1}}
\label{eq:bwd_diff}
\end{align}

We can use (\ref{eq:fwd_diff}) and (\ref{eq:bwd_diff}) to approximate the first derivatives of the displacement components by

\begin{equation}
u'_i \approx \frac{h_lu'^\rightarrow_i + h_ru'^\leftarrow_i}{h_l+h_r}
\label{eq:first_derivative_with_jump_conditions}
\end{equation}

with $h_l = \left(\Delta x_{i-1} + \Delta x_i\right)/2$ and $h_r = \left(\Delta x_i + \Delta x_{i+1}\right)/2$. Note that this is equivalent to the regular 3-point stencil for non-equidistant meshes, if there is no jump in the material parameters. At the boundary nodes, (\ref{eq:fwd_diff}) and (\ref{eq:bwd_diff}) can be used on their own, or, if there are no material jumps within three cells from the boundary, the usual second order approximation

\begin{align}
u'_0 &= \frac{-u_2 h_0^2 + u_1(h_0+h_1)^2 - u_0 h_1(2h_0+h_1)}{h_0 h_1(h_0 + h_1)} \label{eq:gradient_boundary_second_order_0}\\
u'_{N-1} &= \frac{u_{N-3} h_{N-2}^2 - u_{N-2}(h_{N-2}+h_{N-3})^2 + u_{N-1} h_{N-3}(2h_{N-2}+h_{N-3})}{h_{N-2} h_{N-3}(h_{N-2} + h_{N-3})}
\label{eq:gradient_boundary_second_order_N-1}
\end{align}

with $h_i = \left(\Delta x_i + \Delta x_{i+1}\right)/2$ can be used.\\
We employ the first derivative approximation (\ref{eq:first_derivative_with_jump_conditions}) when calculating the mechanical stress and strain, and when calculating the mixed derivatives that enter the force density $\bm{f}$ in (\ref{eq:elastodynamics_equations}). For the second derivatives in $\bm{f}$, we use the approximation

\begin{equation}
\frac{\partial}{\partial x}C\frac{\partial u}{\partial x}\approx C_i\frac{u'^\rightarrow_i - u'^\leftarrow_i}{\Delta x_i}
\label{eq:second_derivative_with_jump_conditions}
\end{equation}

which can be understood by keeping in mind that (\ref{eq:fwd_diff}) and (\ref{eq:bwd_diff}) where obtained from the differences (\ref{eq:u_continuity_l}) and (\ref{eq:u_continuity_r}) from the cell center to the cell boundary. Table~\ref{tab:jump_conditions_lookup} can be used as a reference to populate the two types of jump parameters, $C$ and $B$. The two values of the $B$ parameter cannot be taken at the cell center, however, but need to be the values close at the interface coming the $+$ or $-$ side. For example, if we are interested in derivatives along the $x$-direction, they are

\begin{align}
B^+_i &= \lim_{x\rightarrow x_i+\Delta x_i/2}B(x) \\
B^-_i &= \lim_{x_i-\Delta x_i/2\leftarrow x}B(x).
\end{align}

Herein lies it's own collection of complications: To obtain interface values of the magnetic stress $\sigma^m$, we compute interface values of $\m$ using the proper exchange jump conditions \cite{Heistracher_2022_JMMM}, analogous to (\ref{eq:u_continuity_l}) and (\ref{eq:u_continuity_r}). To compute new interface values of other gradient components, we use the harmonic mean weighted by the corresponding stiffness matrix component. However, this requires prior knowledge of these gradient components in the cell centers, and in consequence, an iterative approach is necessary. In a first step, the gradient components of $\ud$ are calculated without taking other derivatives of $\ud$ into account. The $B$-parameters are then updated based on the computed gradient of $\ud$, and the procedure is repeated.

\begin{table}[H]
\begin{tabular}{cc|ll}
$\partial_{x_i}$ & $u$ & $C$ & \multicolumn{1}{c}{$B$}\\\hline
$\partial_x$ & $u_x$ & $C_{11}$ & $C_{12}\partial_y u_y + C_{13}\partial_z u_z-\sigma^m_{xx}$ \\
$\partial_x$ & $u_y$ & $C_{66}$ & $C_{66}\partial_y u_x-\sigma^m_{xy}$ \\
$\partial_x$ & $u_z$ & $C_{55}$ & $C_{55}\partial_z u_x-\sigma^m_{xz}$ \\&&&\\

$\partial_y$ & $u_x$ & $C_{66}$ & $C_{66}\partial_x u_y-\sigma^m_{xy}$  \\
$\partial_y$ & $u_y$ & $C_{22}$ & $C_{12}\partial_x u_x + C_{23}\partial_z u_z -\sigma^m_{yy}$ \\
$\partial_y$ & $u_z$ & $C_{44}$ & $C_{44}\partial_z u_y-\sigma^m_{yz}$ \\&&&\\

$\partial_z$ & $u_x$ & $C_{55}$ & $C_{55}\partial_x u_z-\sigma^m_{xz}$ \\
$\partial_z$ & $u_y$ & $C_{44}$ & $C_{44}\partial_y u_z-\sigma^m_{yz}$ \\
$\partial_z$ & $u_z$ & $C_{33}$ & $C_{13}\partial_x u_x + C_{23}\partial_y u_y-\sigma^m_{zz}$ \\\hline
\end{tabular}
\caption{Parameter reference for the jump conditions in the spacial derivatives of $\ud$ for materials that can be described by (\ref{eq:stiffness_matrix_covered}). The $C$ and $B$-type parameters are stated for the derivative indicated by the first column and the component of $\ud$ indicated by the second column. }
\label{tab:jump_conditions_lookup}
\end{table}

\subsection{Neumann Boundary Conditions}

From (\ref{eq:LE_weak_from}), we can see that the Neumann boundary conditions of the elastic problem are given by $\left(\sigma
-\sigma^m\right)\cdot\hat{n}=\bm{t}$ at surfaces $\partial\Omega_u^t$. The traction vector $\bm{t}$ is zero at open boundaries, which are the natural boundary conditions of the weak form. To implement these boundary conditions, the boundary values of the stress derivatives parallel to the free surface normal vector $\hat{n}$ in the force density field $\bm{f}=\bm{\nabla}\cdot\left(\sigma-\sigma^m\right)$ need to be updated. In the example of an open yz-plane with $\hat{n}=-\hat{x}$, we would thus need to set the values in the boundary cells located at $x_0$ for the first term in the force field components

\begin{equation}
f_i = \frac{\partial}{\partial x}\left(\sigma_{ix}-\sigma^m_{ix}\right) 
+ \frac{\partial}{\partial y}\left(\sigma_{iy}-\sigma^m_{iy}\right)
+ \frac{\partial}{\partial z}\left(\sigma_{iz}-\sigma^m_{iz}\right).
\label{eq:force_density_components}
\end{equation}

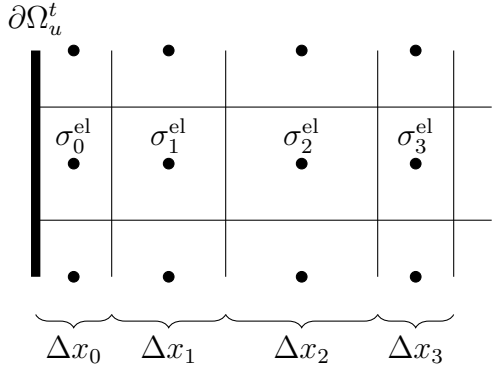
\begin{figure}
\begin{tikzpicture}
\draw[line width=3.5pt] (0,-1.5) -- (0,1.5) node[above] {$\partial\Omega_u^t$}; 
\draw (1,-1.5) -- (1,1.5);
\draw (2.5,-1.5) -- (2.5,1.5);
\draw (4.5,-1.5) -- (4.5,1.5);
\draw (5.5,-1.5) -- (5.5,1.5);
\draw (0,-0.75) -- (6,-0.75);
\draw (0, 0.75) -- (6, 0.75);

\foreach \y in {-1.5, 0, 1.5} {
  \filldraw [black] (0.5, \y) circle (2pt); 
  \filldraw [black] (1.75, \y) circle (2pt);
  \filldraw [black] (3.5, \y) circle (2pt);
  \filldraw [black] (5, \y) circle (2pt);
}

\node[above] at (0.5,0) {$\sigma^\text{el}_0$};
\node[above] at (1.75,0) {$\sigma^\text{el}_1$};
\node[above] at (3.5,0) {$\sigma^\text{el}_2$};
\node[above] at (5,0) {$\sigma^\text{el}_3$};

\draw[decorate,decoration={brace,mirror,amplitude=5pt}]
  (0,-2) -- (1,-2) node[midway,below=4pt] {$\Delta x_0$};

\draw[decorate,decoration={brace,mirror,amplitude=5pt}]
  (1,-2) -- (2.5,-2) node[midway,below=4pt] {$\Delta x_1$};

\draw[decorate,decoration={brace,mirror,amplitude=5pt}]
  (2.5,-2) -- (4.5,-2) node[midway,below=4pt] {$\Delta x_2$};

\draw[decorate,decoration={brace,mirror,amplitude=5pt}]
  (4.5,-2) -- (5.5,-2) node[midway,below=4pt] {$\Delta x_3$};
\end{tikzpicture}
\caption{2D-cut of the irregular finite-differences mesh near the boundary.}
\label{fig:Neumann_BCs}
\end{figure}

Since the degrees of freedom are located at the center of the mesh's cells, the open boundaries do not coincide with the sampling positions. Instead, they are shifted by half of the discretization length, as depicted in figure~\ref{fig:Neumann_BCs}. This allows us to make use of the first derivative scheme for irregular grids 

\begin{equation}
f^{(x)}_{i,0} \approx \frac{\sigma^{el}_{ix,1} h_l^2 - h_r^2 t_x + (h_r^2-h_l^2)\sigma^{el,(2)}_{ix,0}}{h_l^2h_r + h_r^2h_l}
\label{eq:f_Neumann_O2}
\end{equation}

with $h_l=\Delta x_0/2$ and $h_r=\left(\Delta x_0 + \Delta x_1\right)/2$. The boundary value of the stress $\sigma^{el,(2)}_{ix,0}$ was constructed using the second order approximation (\ref{eq:gradient_boundary_second_order_0}). Consequently, we can only use this scheme if there is no jump in the material parameters between the boundary cells and the next two layers of adjacent bulk cells. If (\ref{eq:gradient_boundary_second_order_0}) cannot be used, a different approach would be to use a midpoint formula

\begin{equation}
f^{(x)}_{i,0} \approx \frac{\sigma^{el}_{ix,1} -t_x}{h_l + h_r/2}
\label{eq:f_Neumann_Midpoint}
\end{equation}

where all strain components are calculated from the bulk formula (\ref{eq:first_derivative_with_jump_conditions}). This requires the material to be homogeneous within the first two layers of cells only. If there is a jump in the material parameters after the boundary layer of cells already, one needs to fall back to the forward difference

\begin{equation}
f^{(x)}_{i,0} \approx \frac{\sigma^{el,(1)}_{ix,0} -t_x}{h_l}
\label{eq:f_Neumann_O1}
\end{equation}

with the stress boundary value $\sigma^{el,(1)}_{ix,0}$ computed from the forward difference scheme (\ref{eq:fwd_diff}).

\section{Results}

%
%
%
%
%
%

\subsection{Static Magnetic Strain in an FM/NM Stack}
Let's assume we have a homogeneous nickel film of thickness $L_x$, with one boundary hold in place at $x=0$ and periodic boundary conditions in the $y$ and $z$ directions. First, let's consider the case of there being an open boundary at $x=L_x$. Then the natural boundary condition of the magnetoelastic problem requires that 

\begin{equation}
\sigma\cdot\hat{x}\Big\vert_{x=L_x} = \sigma^m\cdot\hat{x}\Big\vert_{x=L_x}.
\label{eq:natural_bc_strain_example}
\end{equation}

In the static case, the local force density $\bm{f}$ vanishes and thus, the stress in the entire material needs to be constant. One might find it instructive to see how these assumptions already lead to the minimization of the potential energy (\ref{eq:elastic_potential}). Due to the 1D-nature of this problem, the derivatives of $\ud$ along $y$ and $z$ vanish and the change of the displacement components becomes \cite{Shu_2004_MechMater}

\begin{align}
r_x = \frac{\partial u_x}{\partial x} &= \varepsilon^m_{xx} + \frac{C_{12}}{C_{11}}\left(\varepsilon^m_{yy}+\varepsilon^m_{zz}\right) \label{eq:mag_stain_kx}\\
r_y = \frac{\partial u_y}{\partial x} &= 2\varepsilon^m_{xy} \\
r_z = \frac{\partial u_z}{\partial x} &= 2\varepsilon^m_{xz}.
\end{align}

If a homogeneous non-magnetic material of arbitrary thickness is stacked on top of the ferromagnetic material, in the static case, the former boundary condition (\ref{eq:natural_bc_strain_example}) would still hold true, but now assumes the role of a jump condition. This is because no strain is present in the relaxed non-magnetic material, which has an open boundary. The displacement $\ud$ therefore becomes

\begin{equation}
u_i = \left\lbrace\begin{array}{ll}
r_i x & \text{ if } x \leq L_\text{FM} \\
r_i L_\text{FM} & \text{ else }
\end{array} \right..
\label{eq:ud_strain_example}
\end{equation}

\begin{figure}[H]
\includegraphics[width=0.6\textwidth]{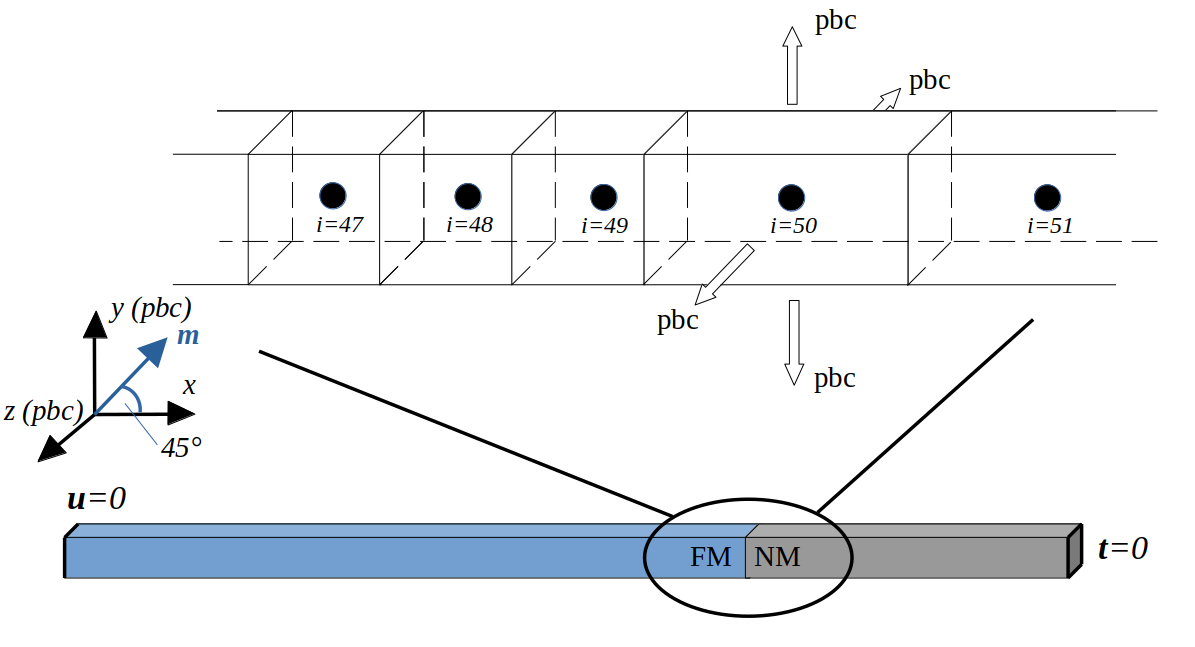}
\caption{Setup of the Ni/Al stack experiment. The stack is sampled by 60 cells, 50 of which are Ni. The cell are cubic in Ni with $\Delta x_\text{Ni}=5\text{ nm}$, while the length of the Al cells is $\Delta x_\text{Al}=10\text{ nm}$. The stack is fixed by homogeneous Dirichlet boundary conditions on the right and the surface at the left is open (homogeneous Neumann boundary conditions). Periodic boundary conditions are applied in the $y$ and $z$-direction.}
\label{fig:1d_stack}
\end{figure}

\begin{figure}[H]
\includegraphics[width=0.75\textwidth]{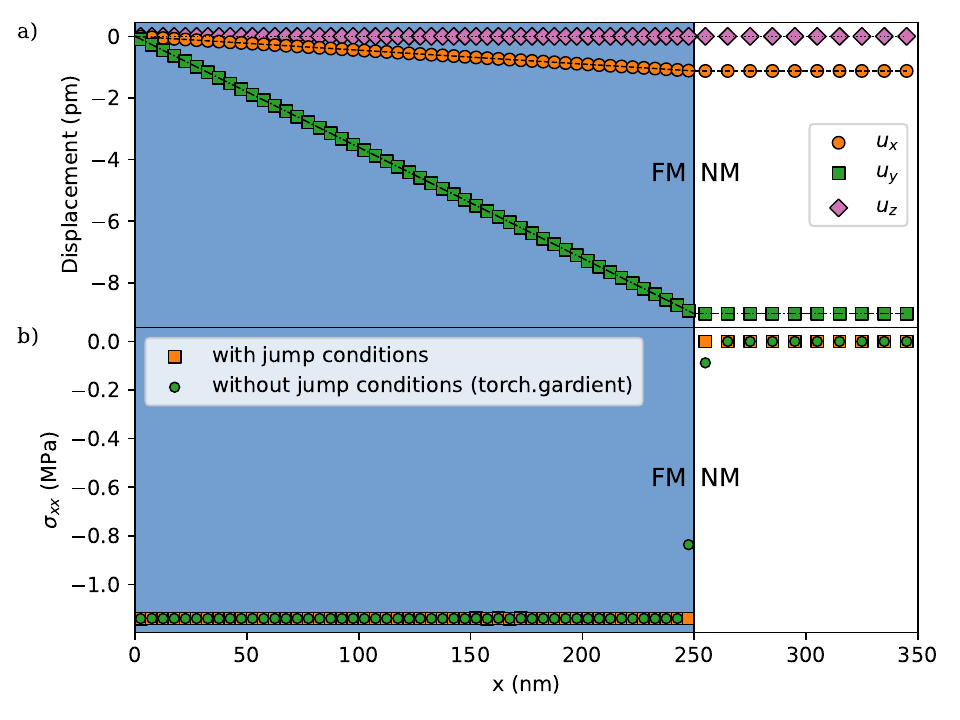}
\caption{a) Relaxed displacement of a Ni/Al stack due to stresses induced by a homogeneous magnetization $\m=(1/\sqrt{2},1/\sqrt{2},0)$. The dashed lines indicate the expected displacement of (\ref{eq:ud_strain_example}), to which the obtained displacement shows excellent agreement. b) The stress component $\sigma_{xx}$ computed from the relaxed $\ud$ by two methods: the first derivative scheme with jump conditions (\ref{eq:first_derivative_with_jump_conditions}) (orange squares) and a 3-point stencil for non-equidistant spacing as implemented by torch.gradient \cite{Paszke_2019_arXiv,pytorch_torch_gradient} (green circles). Only the former correctly approximates $\sigma_{xx}$ as discontinuous over the interface and constant in the two materials. Inside Ni, $(\sigma-\sigma^m)=0$ holds true.}
\label{fig:deformation_from_magnetic_strain}
\end{figure}

Using high values for the Gilbert damping $\alpha$ and the phenomenological elastic damping $\eta$, we relax a stack composed of ${100\text{ nm}}$ of Al on top of ${250\text{ nm}}$ of Ni. The simulation setup is depicted in figure~\ref{fig:1d_stack}. The used mesh is non-equidistant, with a discretization length of $5\text{ nm}$ in Ni and $10\text{ nm}$ in Al, respectively. The elastic properties used for mono-crystalline nickel are $C_{11} = 2.54\times 10^{11}\text{ N/m}^2$, $C_{12} = 1.55\times 10^{11}\text{ N/m}^2$, $C_{44} = 1.23\times 10^{11}\text{ N/m}^2$ \cite{Ledbetter_1973_JPhysChemRefData}, the magnetostrictive coupling parameters are $\lambda_{100} = -4.6\times 10^{-5}$, $\lambda_{111} = -2.4\times 10^{-5}$ \cite{Liang_2014_Nanotechnology}, the exchange constant was set to $A=8\times 10^{-12}\text{ J/m}$ and the saturation magnetization to $M_s=490\text{ kA/m}$. The magnetization is kept along $(1/\sqrt{2},1/\sqrt{2},0)$ by an $200\text{ mT}$ external field. For the aluminium capping, the elastic parameters are $E = 69.6\text{ GPa}$ and $\nu = 0.33$ \cite{Comte_2002_SurfCoatTechnol}.\\ 
The relaxed $\ud$ is shown in figure~\ref{fig:deformation_from_magnetic_strain} a) and due to the inclusion of the jump conditions in the forward and backward differences (\ref{eq:fwd_diff}) and (\ref{eq:bwd_diff}), it matches the expected displacement (\ref{eq:ud_strain_example}). In figure~\ref{fig:deformation_from_magnetic_strain} b), we can compare the stress component $\sigma_{xx}$ computed from the first derivatives including jump conditions (\ref{eq:first_derivative_with_jump_conditions}) to the stress obtained from derivatives computed by the torch.gradient method, which employs a 3-point stencil for non-equidistant grids \cite{Paszke_2019_arXiv,pytorch_torch_gradient}. While the former implementation correctly yields the stress as constants in the two domains, an implementation of the first derivative that does not account for the jump conditions deviates significantly from the expected values at the material interface.  

\subsection{Transversal Bulk Mode Damping}
When an acoustic wave travels through a magnetoelastic material, energy can be transferred to the magnetic system and excite a spin-wave when there is a strain component that leads to torque on the magnetization. Let's assume that the area $0 \leq x \leq L_x$ is occupied by mono-crystalline nickel. The surface at $x=L_x$ is fixed in space and an external shear traction 

\begin{equation}
\bm{t}(t) = \left(0, 0, t_0\sin{\left(\omega t\right)}\right)^\text{T}
\label{eq:bc_t_bulk_wave}
\end{equation}

on the surface at $x=0$ excites a bulk shear wave that travels along the $x$-direction. Due to the form of the boundary condition (\ref{eq:bc_t_bulk_wave}), the only excited strain component is $\varepsilon_{xz}$. Then, the torque acting on the magnetization becomes

\begin{equation}
\bm{T}=6C_{44}\lambda_{111}\varepsilon_{xz}\left(\begin{array}{c}
-m_y m_x  \\ m_x^2-m_z^2 \\ m_y m_z 
\end{array}\right)
\end{equation}

We can see that the excited shear mode will thus be unable to excite a spin wave, when the magnetization is uniformly directed along $\hat{y}$. 
On the other hand, we can see from the magnetoelastic force density $\bm{f}^m=-\nabla\cdot\sigma^m$ that a magnetization precessing around the $z$-direction couples back largely only into the $f^m_z$ force density component. Therefore, a magnetization that is initially directed along $\hat{z}$ is not expected to give rise to a significant $\varepsilon_{xy}$ wave, like it would be the case if $\m$ would precess around the $x$-axis.\\

When there is no contribution to the effective field $\heff$ with a time dependence that does not solely arise due to the time dependence of the magnetization ${\m(t)}$, the dynamics of the magnetization can only lose energy \cite{Abert_2019_EurPhysJB}. Since the magnetoelastic energy depends also on the time-dependent mechanical displacement ${\ud(t)}$, the situation is slightly more involved here. We ignore the mechanical damping, and treat the energy contributions that change in time only because of the magnetization as in reference~\cite{Shu_2004_MechMater}. By substituting $\rho\ddot{\ud}=\bm{\nabla}(\sigma-\sigma^m)$ after applying the time derivative to the kinetic term, we are left only with one additional term that depends on the Neumann boundary conditions.


\begin{align}
\frac{\dtot E}{\dtot t} =& \frac{\dtot}{\dtot t}\int_\Omega \left[\frac{\rho\dot{\ud}^2}{2} + \frac{1}{2}\left(\varepsilon-\varepsilon^m\right):\left(\sigma-\sigma^m\right) - J_s\bm{H}^\text{ext}\cdot\m + A\left(\bm{\nabla}\m\right)^2 \right]\dtot\bm{x} \\
=& \int_\Omega \Big[\rho\dot{\ud}\cdot\ddot{\ud} + \dot{\varepsilon}:\left(\sigma-\sigma^m\right) + \left\lbrace -J_s\bm{H}^\text{ext} - \left(\sigma-\sigma^m\right):\bm{\nabla}_m\varepsilon^m +\bm{\nabla}_m A\left(\bm{\nabla}\m\right)^2\right\rbrace\cdot\dot{\m}  \Big]\dtot\bm{x} \\
=& \int_\Omega \left[\rho\dot{\ud}\cdot\ddot{\ud} + \dot{\varepsilon}:\left(\sigma-\sigma^m\right) -J_s\heff\cdot\dot{\m} \right]\dtot\bm{x} \\
=& \int_\Omega \left[\bm{\nabla}\cdot\left(\sigma-\sigma^m\right)\cdot\dot{\ud} + \dot{\varepsilon}:\left(\sigma-\sigma^m\right) -J_s\heff\cdot\dot{\m}\right]\dtot\bm{x} \\
=& \int_\Omega \left[ \dot{\varepsilon}:\left(\sigma-\sigma^m\right) - \left(\sigma-\sigma^m\right):\dot{\varepsilon} - J_s\heff\cdot\dot{\m}  \right]\dtot\bm{x} + \int_{\partial\Omega}\bm{t}\cdot\dot{\ud}\,\dtot \bm{S} \\
=& -\int_\Omega J_s\heff\cdot\dot{\m}\,\dtot\bm{x} + \int_{\partial\Omega}\bm{t}\cdot\dot{\ud}\,\dtot \bm{S} \\
=& -\int_\Omega J_s\frac{\alpha\gamma}{1+\alpha^2}\left\vert\m\times\heff\right\vert^2\,\dtot\bm{x} + \int_{\partial\Omega}\bm{t}\cdot\dot{\ud}\,\dtot \bm{S} \label{eq:gain_and_loss}
\end{align}

\begin{figure}[H]
\includegraphics[width=0.7\textwidth]{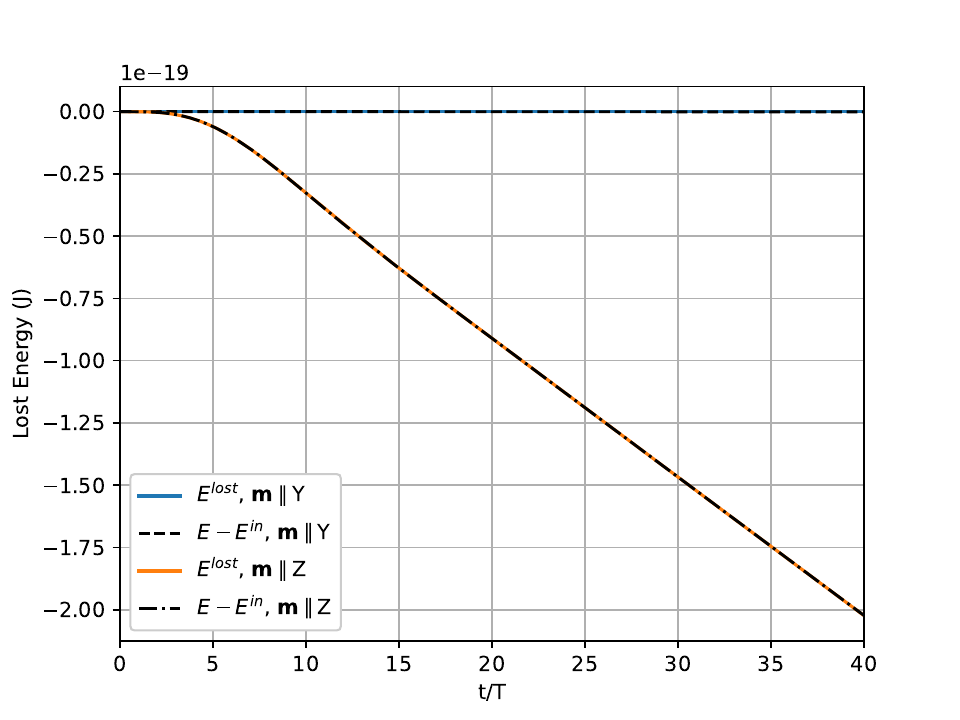}
\caption{Gilbert losses in the bulk shear wave setup. No energy is lost for $\bm{H}^\text{ext}\parallel\hat{y}$ since no excitation of an SW occurs. The system is otherwise undamped ($\eta=0$), so that the losses observed for $\bm{H}^\text{ext}\parallel\hat{z}$ are equal to the difference between the total free energy and the energy injected by the Neumann boundary conditions.}
\label{fig:bulk_energy_loss}
\end{figure}

\begin{figure}[H]
\includegraphics[width=0.7\textwidth]{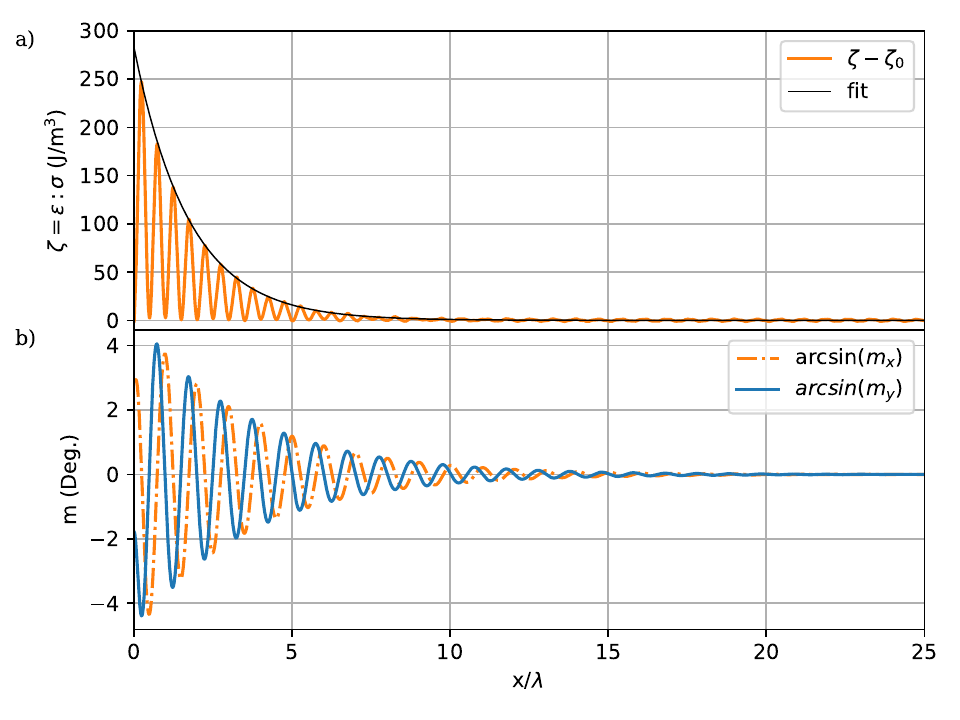}
\caption{Attenuation of a shear bulk mode ($\varepsilon_{xz}\neq 0$) propagating in $x$-direction in Ni for $\bm{H}^\text{ext}\parallel\hat{z}$. a) Exponential energy loss along the propagation direction. No components other than $\varepsilon_{xz}$ contribute significantly to the potential energy. The exponential damping parameter obtained from a fit is $-1.268\times 10^{-6}\text{ m}^{-1}$. b) The SW amplitude is only large near the open boundary, where the amplitude of the acoustic shear wave is still high.}
\label{fig:bulk_damping_z}
\end{figure}

This additional boundary term vanishes for open boundary conditions ($\bm{t}=0$). In the presented experiment however, this term yields the energy injected into the system. \\
The initial displacement is obtained analogous to (\ref{eq:mag_stain_kx}), but under the assumption that the right boundary is fixed. The anisotropy and stray field are neglected in this setup, and therefore, the resonance field can be computed from 

\begin{equation}
h^\text{ext} = \frac{2\pi f}{\gamma} - \frac{2A}{M_s \mu_0}\left(\frac{2\pi f}{c_t}\right)^2
\end{equation}

where $c_t\approx 3718\text{ m/s}$ is the speed of transversal bulk modes in mono-crystalline nickel. The simulations are conducted at $f=4\text{ GHz}$. The cell lengths were chosen such that a wavelength of the bulk-shear wave is sampled by $200$ cells in the propagation direction ($x$-direction).  Periodic boundary conditions are set along $\hat{y}$ and $\hat{z}$ with only one cell in the transverse directions. For numerical stability, the maximum time step $\Delta t$ is set to satisfy the CFL condition $c_t\Delta t/\Delta x \leq 1$. Note that this is strictly speaking no stability criterion here, since we use the RKF45 scheme with adaptive time steps for error control.\\
The lost energy for the external field and initial magnetization along $\hat{y}$ and $\hat{z}$ are shown in figure~\ref{fig:bulk_energy_loss}. We can see that indeed, having the initial magnetization and the external field along $\hat{y}$ won't lead to any losses, as no coupling to the magnetic system occurs. Having them directed along $\hat{z}$ however allows the shear mode to couple to a spin-wave mode and power is lost due to Gilbert damping. The Gilbert losses, given by the first term in (\ref{eq:gain_and_loss}), agree perfectly with the difference between the total free energy of the system and the injected energy by the boundary condition, obtained from the second term in (\ref{eq:gain_and_loss}). The potential energy density and the dynamic components of the magnetization after 40 periods $T$ are plotted against the propagation length for the first 25 wavelengths in figure~\ref{fig:bulk_damping_z} and the potential energy density exhibits the expected exponential loss~\cite{Dreher_2012_PhysRevB}.


\begin{figure}[H]
\includegraphics[width=0.7\textwidth]{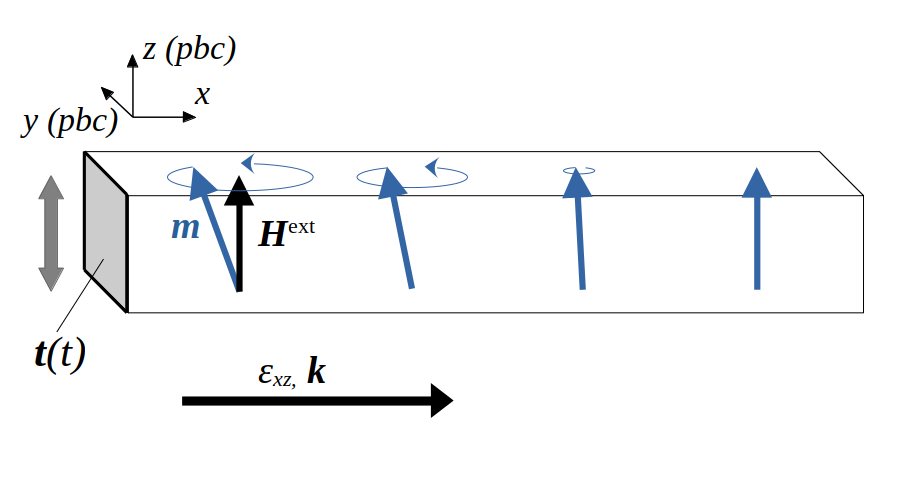}
\caption{A horizontal shear mode with only $\varepsilon_{xz}\neq 0$ is injected by Neumann boundary conditions from the left. For $\bm{H}^\text{ext}\parallel\hat{z}$, the acoustic wave excites a spin wave. Periodic boundary conditions are set in the directions perpendicular to the progpagation direction ($\hat{y}$ and $\hat{z}$).}
\label{fig:setup_bulk_mode}
\end{figure}

\subsection{Boundary conditions for Rayleigh modes}
Before we discuss the interaction of SAWs with SWs in layered systems, we will first direct our attention to the numerical errors of some possible implementations of the open boundary conditions for Rayleigh modes in a non-magnetic substrate. According to reference~\cite{Maekawa_1976_AIP}, the Rayleigh eigenmode propagating in $x$-direction in a homogeneous isotropic substrate filling the half-space in $z>0$ is given by

\begin{align}
u_x(x,z,t) &= A_0\left[\exp{(-\kappa_t z)}-2\frac{k^2}{k^2+\kappa_t^2}\exp{(-\kappa_l z)} \right]\cos{(kx-\omega t)} \label{eq:R_ux}\\
u_z(x,z,t) &= -\frac{A_0 k}{\kappa_t}\left[\exp{(-\kappa_t z)}-2\frac{\kappa_t \kappa_l}{k^2+\kappa_t^2}\exp{(-\kappa_l z)} \right]\sin{(kx-\omega t)} \label{eq:R_uz}
\end{align}

with

\begin{align}
\kappa_l &= \sqrt{k^2-(\omega/c_l)^2} \\
\kappa_t &= \sqrt{k^2-(\omega/c_t)^2}.
\end{align}

The propagation speed $c$ of the Rayleigh mode is related to the longitudinal and transversal velocities, $c_l$ and and $c_t$, by \cite{Landau_1970_TheoryOfElasticity} 

\begin{equation}
\left(\frac{c}{c_t}\right)^6 - 8\left(\frac{c}{c_t}\right)^4 + 8\left(\frac{c}{c_t}\right)^2\left(3-2\frac{c_t^2}{c_l^2}\right) - 16\left(1-\frac{c_t^2}{c_l^2}\right) = 0.
\label{eq:R_comp_condition}
\end{equation}

To approximate the Rayleigh mode in the piezoelectric $36^\circ$-YX cut $\text{LiTaO}_3$ substrate used in reference~\cite{Kuess_2021_PhysRevAppl}, on which we will base our R-SAW and SW interaction studies in the next section, (\ref{eq:R_ux}) and (\ref{eq:R_uz}) were fitted to the open-surface eigenmode obtained from COMSOL Multiphysics \cite{COMSOL} within $\lambda_\text{SAW,R}/5$ of the surface and with (\ref{eq:R_comp_condition}) added as a strongly weighted additional residuum. We do not use the COMSOL eigenmode directly, since the trigonal symmetry of LiTaO$_3$ cannot be described by (\ref{eq:stiffness_matrix_covered}) and therefore, we cannot expect that the jump to the magnetic material would be handled correctly by our gradient implementation. The obtained fit parameters are $c \approx 3104\text{ m/s}$, $c_t \approx 3405\text{ m/s}$, and $c_l \approx 5587\text{ m/s}$. The mass density of the substrate is $\rho = 7450\text{ kg/m}^3$ \cite{Ye_2022_ECSJSolStateSciTech}.\\

\begin{figure}
\includegraphics[width=0.7\textwidth]{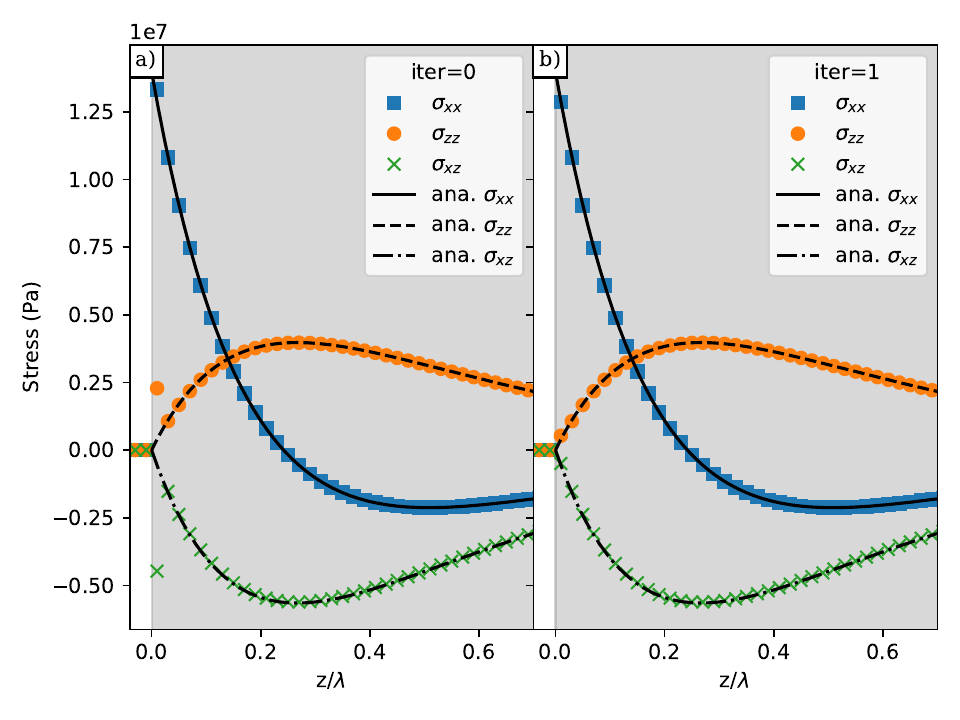}
\caption{Stress components of a Rayleigh mode in an jump-to-air simulation. a) The jump conditions at the interface account only for a jump in $C_{ij}$ to 0, i.e. $\partial_z u_i=0\Big\lvert_{z=0}$. b) The full jump conditions according to table~\ref{tab:jump_conditions_lookup} are used. The strains computed in a) are re-used to build up the $B$-column in table~\ref{tab:jump_conditions_lookup}.}
\label{fig:iter_comp_sig_R_mode}
\end{figure}
\begin{figure}
\includegraphics[width=0.7\textwidth]{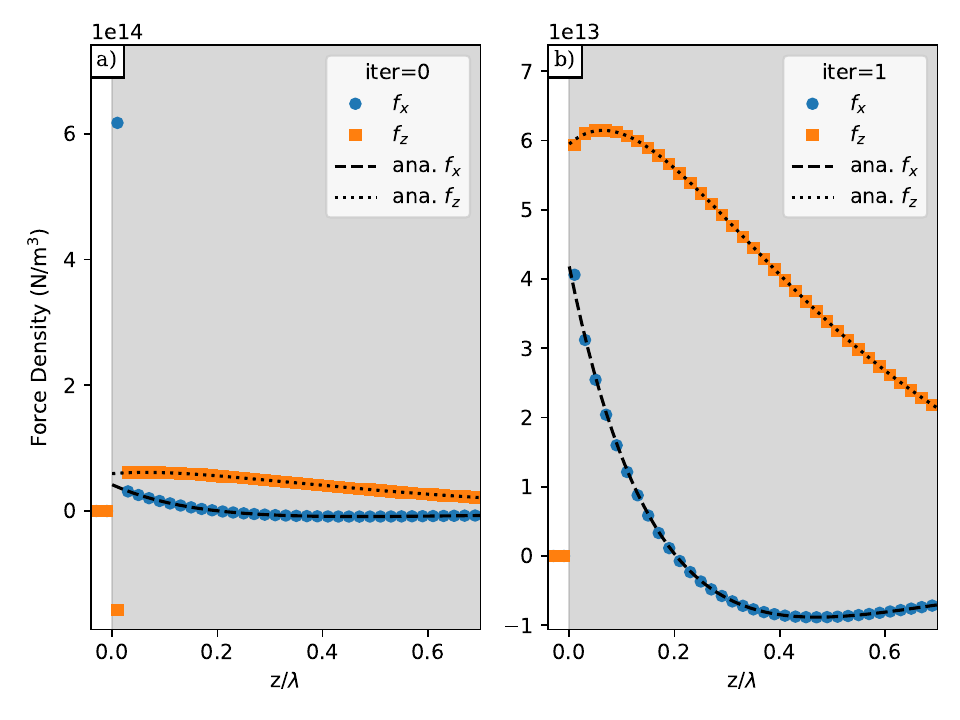}
\caption{Analogous to figure~\ref{fig:iter_comp_sig_R_mode}: The resulting forces computed a) ignoring the $B$-column in table~\ref{tab:jump_conditions_lookup}, and b) using an iterative step to build up and use the $B$-column in the look-up table.}
\label{fig:iter_comp_f_R_mode}
\end{figure}

The values of $\bm{f}$ at the boundary nodes can be obtained from boundary value formulas including 3 boundary nodes (\ref{eq:f_Neumann_O2}), a mixed approach using the bulk formula for the strain computation and the midpoint (\ref{eq:f_Neumann_Midpoint}), or forward difference formulas (\ref{eq:f_Neumann_O1}) only.\\
Alternatively, within the proposed algorithm, it is also possible to introduce additional layers of air ($C^\text{air}_{ij}=0$) to the simulation box. The jump conditions in (\ref{eq:fwd_diff}) and (\ref{eq:bwd_diff}), and in extension in (\ref{eq:first_derivative_with_jump_conditions}) and (\ref{eq:second_derivative_with_jump_conditions}), are then expected to correctly reproduce the open boundary conditions. The jump conditions that enter the differentiation formulas (listed in table \ref{tab:jump_conditions_lookup}) relate the gradient components of $\ud$ to each other, and an iterative step needs to be performed to account for this. In the first iteration, only the $C$-type jump conditions are considered, so that the $B$-type conditions can be built up for a second iteration of the gradient computation. The airbox approach can only approximate the open-boundary conditions successfully with this iterative approach. Note that the boundary value formulas for the force densities are stated in terms of the stress components, and the relations of the relevant strain components $\varepsilon_{xx}$, $\varepsilon_{zz}$, and $\varepsilon_{xz}$, that pose the jump conditions in the airbox approach, can be inferred from them. A comparison of the computed stresses and force densities for $50$ nodes per wavelength at $f=4.47\text{ GHz}$ is depicted in figures~\ref{fig:iter_comp_sig_R_mode} and \ref{fig:iter_comp_f_R_mode}, respectively. Since the gradient components of $\ud$ required to build up the jump conditions do not have jumps in them themselves, further iteration steps do not provide any benefits.\\

\begin{figure}
\includegraphics[width=0.7\textwidth]{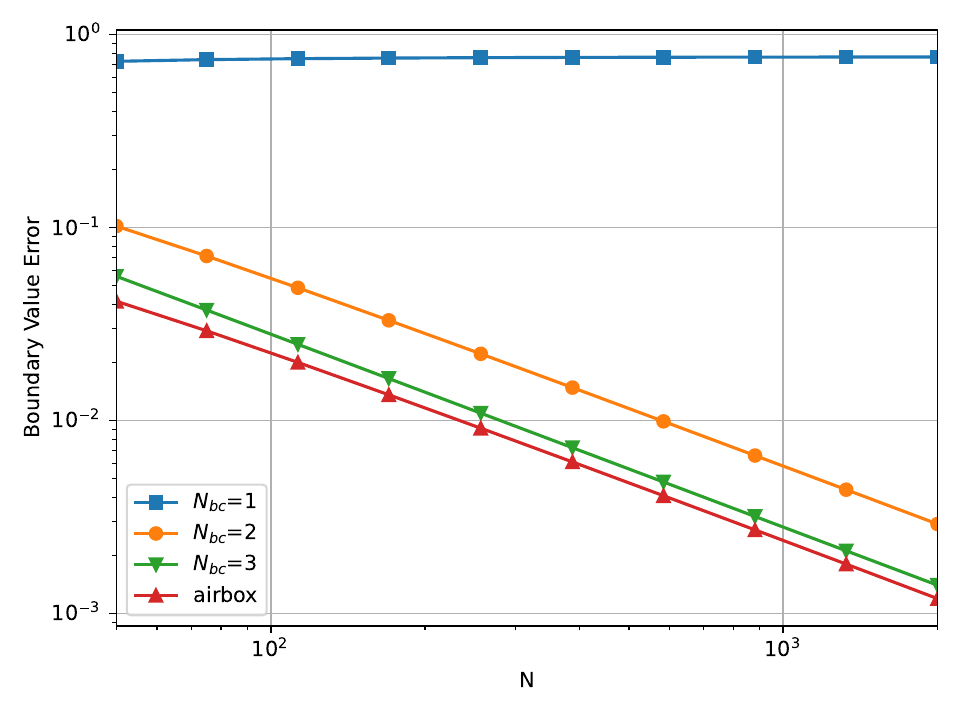}
\caption{Comparison of the relative error in the natural boundary conditions for a Rayleigh-mode simulation using either one of the proposed schemes for Neumann boundary conditions (\ref{eq:f_Neumann_O2}) (${N_\text{bc}=3}$), (\ref{eq:f_Neumann_Midpoint}) (${N_\text{bc}=2}$) or (\ref{eq:f_Neumann_O1}) (${N_\text{bc}=1}$), or with an additional layer of air ($C_{ij}=0$) above the free surface.}
\label{fig:open_boundary_error}
\end{figure}

A comparison of the error for the airbox-method and the boundary value formulas is depicted in figure~\ref{fig:open_boundary_error}. The airbox method is found to yield the most accurate result, while the forward difference method produces an error of order zero, rendering it impractical for the surface acoustic wave simulations. In practice, it can still be used for ferromagnetic free surfaces, where a high damping through the magnetoelastic coupling suppresses high frequency modes that arise from numerical errors. For the airbox method and the three point stencil method, a sampling of around $200$ to $300$ nodes per wavelength will result in an relative error below $10^{-2}$.

\subsection{Rayleigh mode damping in magnetic thin films}
\begin{figure}[H]
\includegraphics[width=0.7\textwidth]{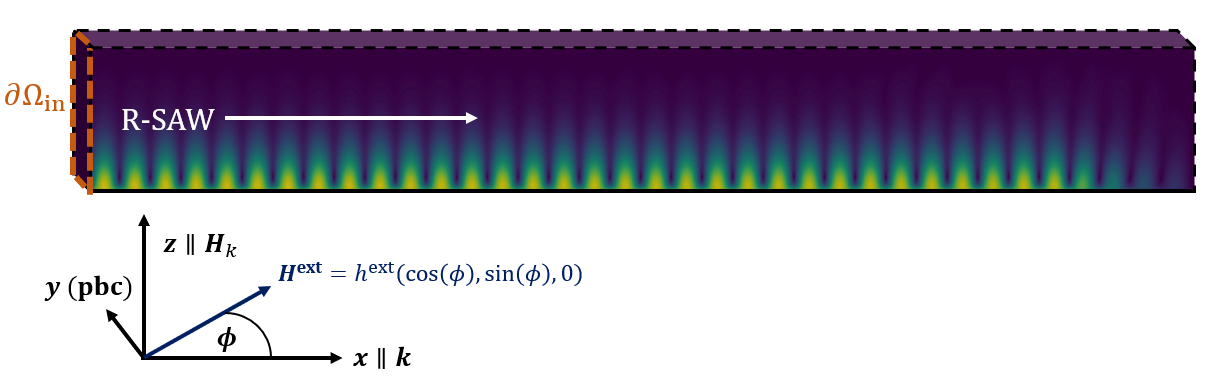}
\caption{Simulation setup for R-SAW damping in an magnetic thin film structure. The open surface is at $z=0$, the substrate eigenmode is injected from the surface at $x=0$ by time-dependent Dirichlet boundary conditions. The surfaces at $x=18\lambda$ and $z=2.3\lambda$ are fixed, PBCs are set in the y-direction. An external field is applied in the $xy$-plane.}
\label{fig:R_setup}
\end{figure}

We now investigate the damping of a Rayleigh SAW in a magnetic thin-film structure depicted in figure~\ref{fig:R_setup}, based on the experimental setup described by Küß et al. in reference~\cite{Kuess_2021_PhysRevAppl}. In their work, they examined how the attenuation of SAWs propagating through a Ni thin film sputtered onto a LiTaO$_3$ substrate depends on both the strength and the orientation of an in-plane bias field ${\bm{H}^\text{ext}=h^\text{ext}(\cos(\phi),\sin(\phi),0)}$, where $\phi$ is the angle between the propagation direction of the SAW and the external field. For the Rayleigh mode, a four-folded symmetry in the power loss was observed, showing the expected asymmetry relative to the external field direction \cite{Dreher_2012_PhysRevB}. Since this asymmetry is between the configurations in which the bias field is pointing to the horizontal left and right of the wave's travel direction, the same asymmetry would be observed when reversing the travel direction while keeping the external field direction. For this reason, the effect is commonly referred to as ``non-reciprocity". \\
Just as in the previous simulation of bulk waves coupling to the magnetic system, we assume that the system will only lose energy through the Gilbert damping ($\eta=0$). Consequently, the non-reciprocity is only due to an asymmetry in the coupling efficiency under reversal of the external field, arising due to a matching or mismatching precessional sense of the SAW's strain and the magnetization dynamics \cite{Sasaki_2017_PhysRevB,Yamamoto_2020_JPhysSocJpn,Xu_2020_SciAdv}.

\subsubsection{Dispersion}
The coupling of an SAW to a SW will only occur around the intersection of the SAW dispersion $2\pi f=ck$ with the SW dispersion. The latter can be obtained from the Sinc-pulse method \cite{Venkat_2013_IEEETransMagn}. We use the same magnetic parameters as in reference~\cite{Kuess_2021_PhysRevAppl} for the polycrystalline Ni thin-film ($M_s = 408\text{ kA/m}$, $A = 7.7 \text{ pJ/m}$) but we omit the rather small uniaxial anisotropy. The value of the in-plane surface anisotropy $H_k$ was varied to achieve the resonant field of $h^\text{ext}=46.5\text{ mT}$ that was observed in the experiment. We find an intersection of the SW and SAW dispersion at $f=4.47\text{ GHz}$ for $H_k=93.1\text{ kA/m}$. The deviation from the fitted value of $158.2\text{ kA/m}$ in reference~\cite{Kuess_2021_PhysRevAppl} is explained by the stray-field approximation \cite{Kalinikos_1986_JPhysCSolStatePhys} used in the source, which we implemented for comparison to validate this claim. For our simulations, we  compute the demagnetization field (\ref{eq:demag_field}) from a numerical FFT-based scheme \cite{Bruckner_2021_SciRep} with periodic boundary conditions in $y$ instead. The repetition number for the periodic boundary conditions in the stray field is chosen such that the film has an assumed equal length in $x$ and $y$. 

\begin{figure}[H]
\includegraphics[width=0.7\textwidth]{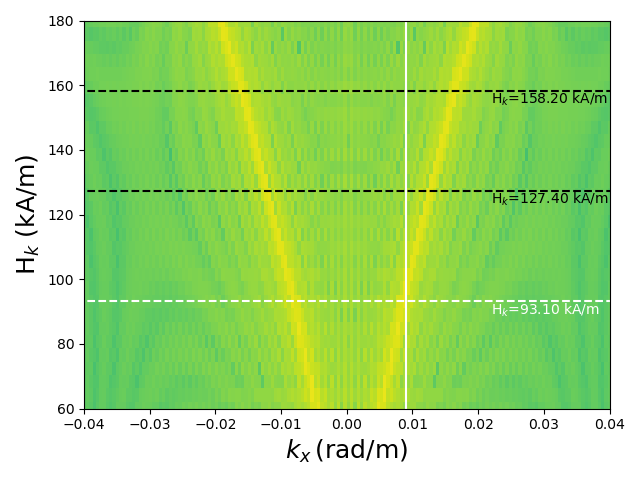}
\caption{Variation of the in-plane anisotropy field to obtain an intersection between the polycrystalline Ni dispersion with the Rayleigh mode for $c=3014\text{ m/s}$ at $f=4.47\text{ GHz}$ and $h^\text{ext}=46.5\text{ mT}$. The obtained value of $93.1\text{ kA/m}$ differs from the fit parameter value of $158.2\text{ kA/m}$ in reference~\cite{Kuess_2021_PhysRevAppl} due to a different stray field approximation and is closer to the value of $127.4\text{ kA/m}$ obtained from a FMR measurement in the same reference. Both values are indicated by black dashed lines.}
\label{fig:Dispersion_HK_sweep}
\end{figure}

\subsubsection{Non-reciprocity}
\begin{figure}[H]
\includegraphics[width=0.7\textwidth]{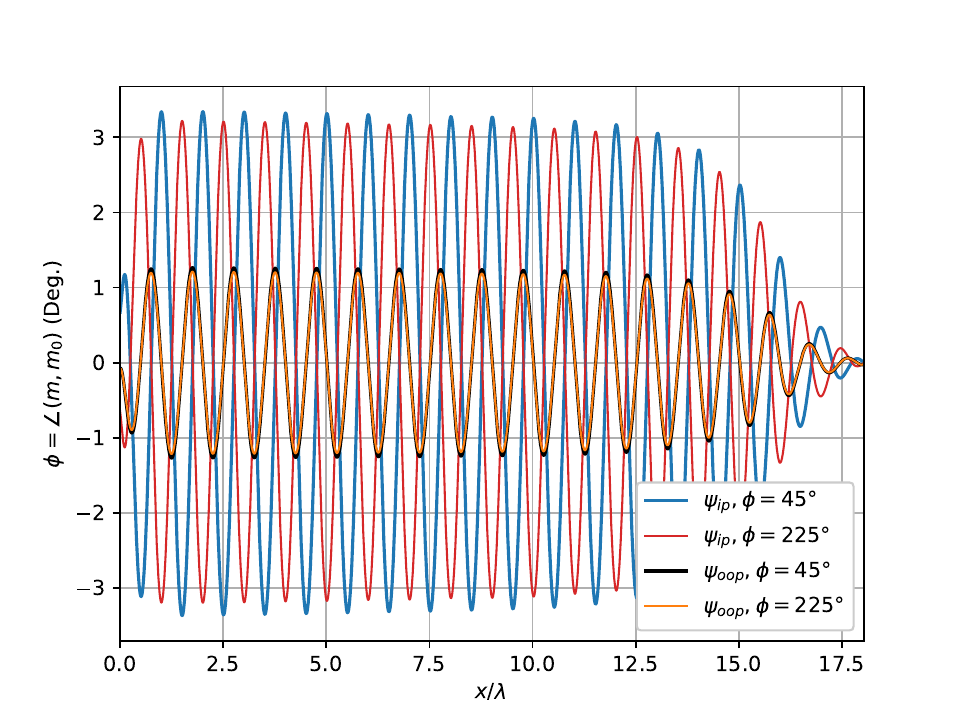}
\caption{SW excited by a R-SAW with $f=4.47\text{ GHz}$ and $h^\text{ext}=46.5\text{ mT}$. The in-plane $\psi_{ip}$ and out-of-plane $\psi_{oop}$ precession angle of the magnetization are shown for an external field angle of $\phi=45^\circ$ and $\phi=225^\circ$. The SW excitation appears slightly larger at $\phi=45^\circ$ because of the helicity mismatch effect \cite{Sasaki_2017_PhysRevB,Yamamoto_2020_JPhysSocJpn,Xu_2020_SciAdv}.}
\label{fig:R_m_signal}
\end{figure}

\begin{figure}[H]
\includegraphics[width=0.7\textwidth]{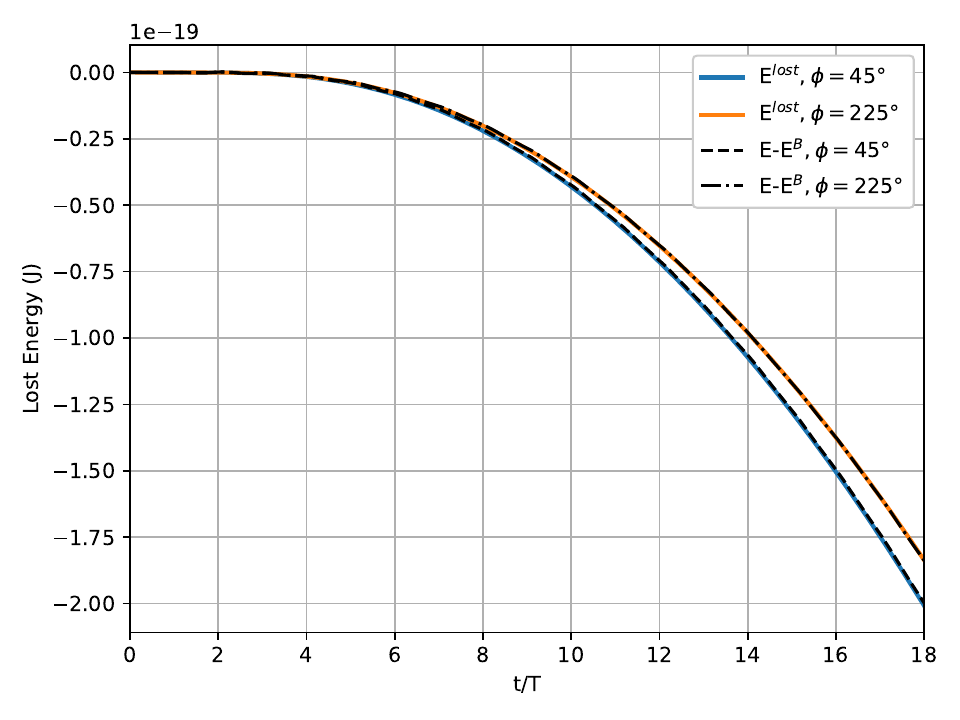}
\caption{Lost energy due to R-SAW and SW coupling, computed from the Gilbert losses and from the difference of the total free energy in the in-resonance simulation and a background simulation at $h^\text{ext}=200\text{ mT}$ and $\phi=0$.}
\label{fig:R_mode_loss_vs_t}
\end{figure}

\begin{figure}[H]
\includegraphics[width=0.7\textwidth]{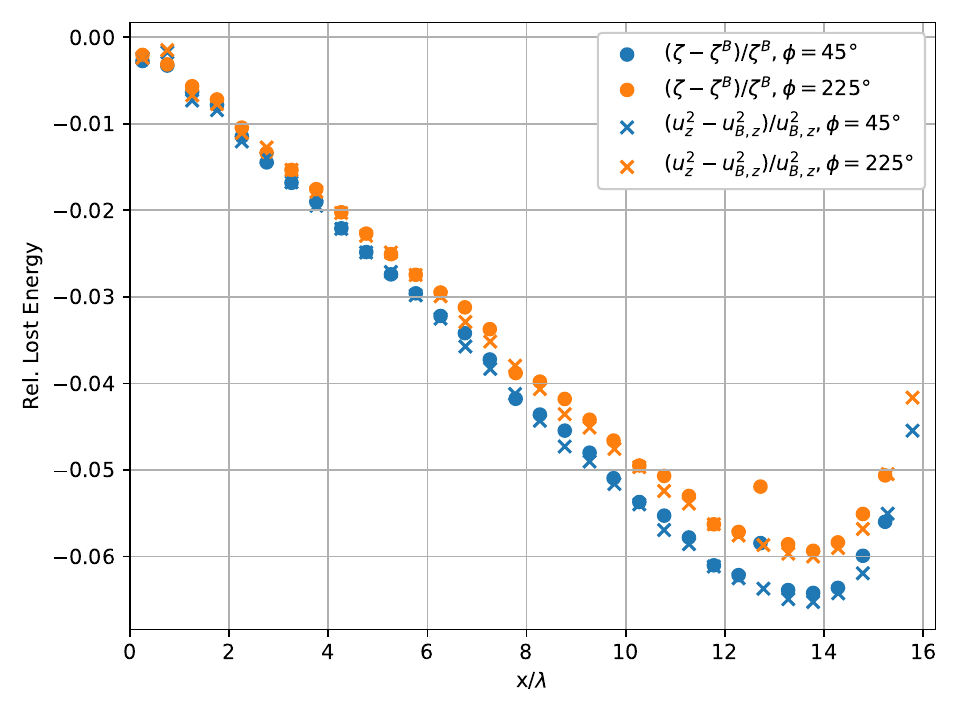}
\caption{Estimation of the R-SAW attenuation from the energy density and the $u_z$ amplitude at mid-height of the Ni-film, plotted against the propagation distance. The peaks to the right appear less damped since the magnetization is only fully excited after a few periods of the R-SAW have passed through.}
\label{fig:R_mode_loss_vs_x}
\end{figure}

At the resonant field determined from the magnetic dispersion, the strength of the coupling between the SAW and the SW depends on the angle $\phi$ between the external field and the propagation direction ($x$-direction). Since we omitted the small unidirectional anisotropy, the maxima of the transmission losses are expected at multiples of $45^\circ$ \cite{Kuess_2021_PhysRevAppl}. The magnetic properties of the Ni thin-film ($M_s = 408\text{ kA/m}$, $A = 7.7 \text{ pJ/m}$, $\alpha = 0.069$) are set in the first three layers of FD-cells $3.47\text{ nm}\times 5\text{ nm}\times 3.\dot{3}\text{ nm}$ from the surface, with the height of this magnetic system being $L_z^\text{FM}=10\text{ nm}$. There are $200$ cells per wavelength $\lambda_\text{SAW,R}$ in the $x$-direction, and only one cell in the $y$-direction with periodic boundary conditions. In the stray field calculation, $1249$ repetitions in $y$ are used, so that the film with $L_x=18\lambda_\text{SAW,R}$ appears square. Since we do not expect the thin film to alter the mechanics drastically, the mechanical material parameters are left unchanged to the parameters found in the wave profile fitting-procedure everywhere. Therefor, (\ref{eq:R_ux}) and (\ref{eq:R_uz}) are still close to an eigenmode of the system. The magnetostrictive coupling constants are set to $\lambda_{100}=\lambda_{111}=-1.89\times 10^{-5}$ within the magnetic region, which equals to a value of $b_1=b_2=12\text{ T}$ for the magnetoelastic coupling constants used in reference~\cite{Kuess_2021_PhysRevAppl}. This value is only a rough estimate, since the actual value of $b_1$ in reference~\cite{Kuess_2021_PhysRevAppl} cannot be determined without knowledge of the proportionality factor between the SAW power and amplitude \cite{Kuess_2020_PhysRevLett}.\\
In order to inject the Rayleigh mode into the system, after fully relaxing the system, the amplitude is ramped up to $A_0 = 10\text{ pm}$ within the first two oscillations using an exponential envelope. By doing so, a band of higher frequencies is excited along with the intended mode. These disturbances will result in large errors when using only forward differences for the boundary value calculations at the open surface and at off-resonance fields, even when a significant phenomenological damping $\eta$ is introduced. Instead, we use the three point stencil method (\ref{eq:f_Neumann_O2}), and no additional damping ($\eta=0$).\\
Figure~\ref{fig:R_m_signal} shows the state of the magnetization at mid-height of the magnetic thin-film at $t=18T$, when the first wavefront of the SAW has reached the end of the simulation box. The in-plane and out-of-plane amplitudes of the magnetization vector from the relaxed position appear larger for the external field angle of $\phi=45^\circ$ than for $\phi=225^\circ$. Conversely, the total lost energy by the system via the Gilbert damping is larger, as seen in figure~\ref{fig:R_mode_loss_vs_t}. The energy loss computed from 

\begin{equation}
E^\text{tot. loss}_\text{m}(t') = -\int_0^{t'}\int_\Omega J_s\frac{\alpha\gamma}{1+\alpha^2}\left\vert\m\times\heff\right\vert^2\,\dtot\bm{x}\dtot t
\end{equation}

matches the free energy difference to a background measurement at $h^\text{ext}=200\text{ mT}$ and $\phi=0^\circ$.  When comparing the angular dependence to the source reference, the reader shall keep in mind that the experimental setup is flipped on its head. Thus, the angles $\phi$ are inverted compared to reference~\cite{Kuess_2021_PhysRevAppl}.\\
A comparison to this background simulation also allows us to investigate the SAW amplitude attenuation and the related potential energy density change, which is depicted in figure~\ref{fig:R_mode_loss_vs_x}. While the displacement data can be used directly to compare the maxima of the resonant and background simulations, high frequency contributions ($>50f$) that arise from numerical errors as well as the injection method were filtered from the energy landscape before the maxima where determined. Both methods agree well with each other, indicating that the overall shape of the SAW is not altered drastically by the magnetoelastic coupling. Due to the simulation box length, the data points in figure~\ref{fig:R_mode_loss_vs_x} do not allow to infer that the attenuation exhibits the expected form of an exponential decay. However, the obtained power losses approximate the experimentally observed attenuation reasonably well. For an isotropic FM layer, in which $\lambda_{100}=\lambda_{111}$, the contrast of the non-reciprocity is determined by the ratio between $\varepsilon_{xz}$ and $\varepsilon_{xx}$ \cite{Sasaki_2017_PhysRevB,Kuess_2020_PhysRevLett}. In the eigenmode obtained from the magnetoelastically decoupled fit, this ratio is $\varepsilon_{xz}/\varepsilon_{xx}=4.30\text{\%}$ at mid-height of the Ni-film ($5\text{ nm}$) and $4.36\text{\%}$ at average. The non-reciprocity is thus found to be slightly smaller than the one observed in reference~\cite{Kuess_2021_PhysRevAppl}, where the strain ratio was $4.95\text{\%}$ and the relative lost power after 10 wavelengths ($\approx 6740\text{ nm}$) would be approximately $4.55\%$ and $3.66\%$ for the high and low damping configurations respectively. 

\subsection{Horizontal shear mode damping in magnetic thick films}
\subsubsection{Eigenmode of the layerd system}
\begin{figure}[H]
\includegraphics[width=0.7\textwidth]{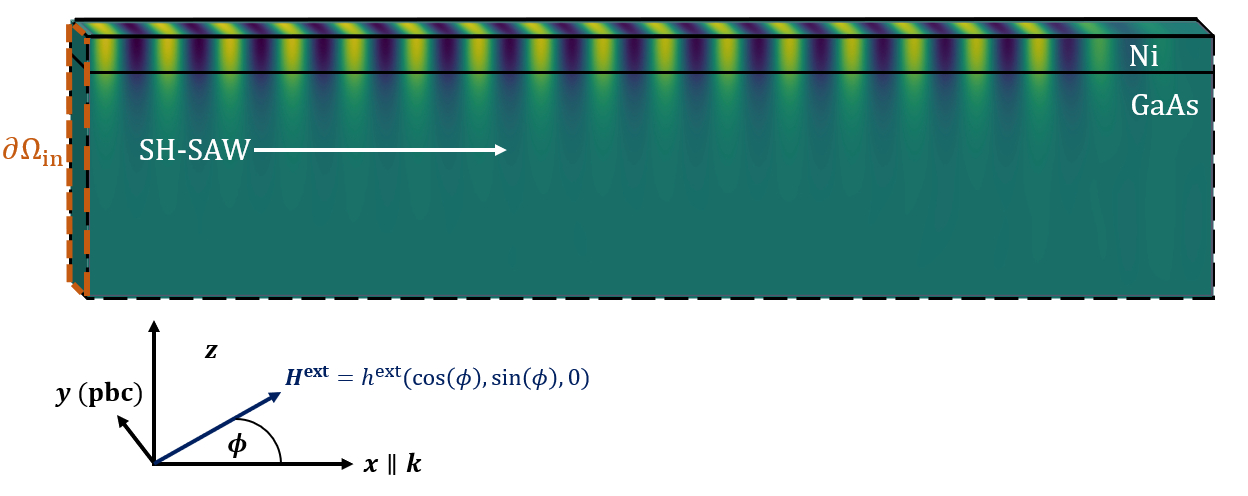}
\caption{Simulation setup for SH-SAW damping in a magnetic thick film structure. The open surface is at $z=300\text{ nm}$, the FM/NM interface is located at $z=0$. The eigenmode of the layered system is injected from the surface at $x=0$ by time dependent Dirichlet boundary conditions. The surfaces at $x=18\lambda$ and $z=-3.5\lambda$ are fixed, PBCs are set in the y-direction. An external field is applied in the $xy$-plane.}
\label{fig:SH_setup}
\end{figure}

Finally, we will look at a layered system of $300\text{ nm}$ of polycrystalline Ni on top of a GaAs substrate. The simulation setup is shown in figure~\ref{fig:SH_setup}. The elastic parameters for GaAs are taken from \cite{Azovtsev_2021_PhysRevMater} and are $C^\text{GaAs}_{11}=1.188\times10^{11}\text{ N/m}^2$, $C^\text{GaAs}_{12}=5.37\times10^{10}\text{ N/m}^2$, $C^\text{GaAs}_{44}=5.94\times10^{10}\text{ N/m}^2$, $\rho^\text{GaAs}=5317\text{ kg/m}^3$. Compared to the setup of the Rayleigh mode experiment earlier, a significant part of the SAW is located in the Ni-film now, and we thus account for the elastic properties of the Ni-film ($E=218\times 10^{9}$, $\nu=0.3$, $\rho=8900\text{ kg/m}^3$ \cite{Ledbetter_1973_JPhysChemRefData,Kuess_2021_PhysRevAppl}). The magnetostrictive coupling parameters are $\lambda_{100}=\lambda_{111}=-1.95\times 10^{-5}$.  According to \cite{Landau_1970_TheoryOfElasticity}, in a system layered along $\hat{z}$, the horizontal shear mode propagating in the direction of $\hat{x}$ exhibits only a $y$-component and can be parametrized by 

\begin{equation}
u_{y}=\left\lbrace\begin{array}{ll}
\left(A_0\cos{\left(z\kappa^\text{FM}\right)} + B\sin{\left(z\kappa^\text{FM}\right)}\right)\sin{\left(kx-\omega t\right)}  & \text{if } z\ge 0\\
A_0\exp{\left(z\kappa^\text{NM}\right)}\sin{\left(kx-\omega t\right)} & \text{else}
\end{array}\right.
\end{equation}

with

\begin{equation}
B = A_0\frac{\kappa^\text{NM}C^\text{NM}_{44}}{\kappa^\text{FM}C^\text{FM}_{44}}
\end{equation}

and 

\begin{align}
\kappa^\text{FM}&=\sqrt{\frac{\omega^2}{c^\text{FM}_t}-k^2} \\
\kappa^\text{NM}&=\sqrt{k^2-\frac{\omega^2}{c^\text{NM}_t}}.
\end{align}

From reference~\cite{Landau_1970_TheoryOfElasticity}, we learn also that the compatibility condition is

\begin{equation}
\tan{\left(h\kappa^\text{FM}\right)} = \frac{\kappa^\text{NM}C^\text{NM}_{44}}{\kappa^\text{FM}C^\text{FM}_{44}} .
\label{eq:SH_compatibility}
\end{equation}

\begin{figure}[H]
\includegraphics[width=0.7\textwidth]{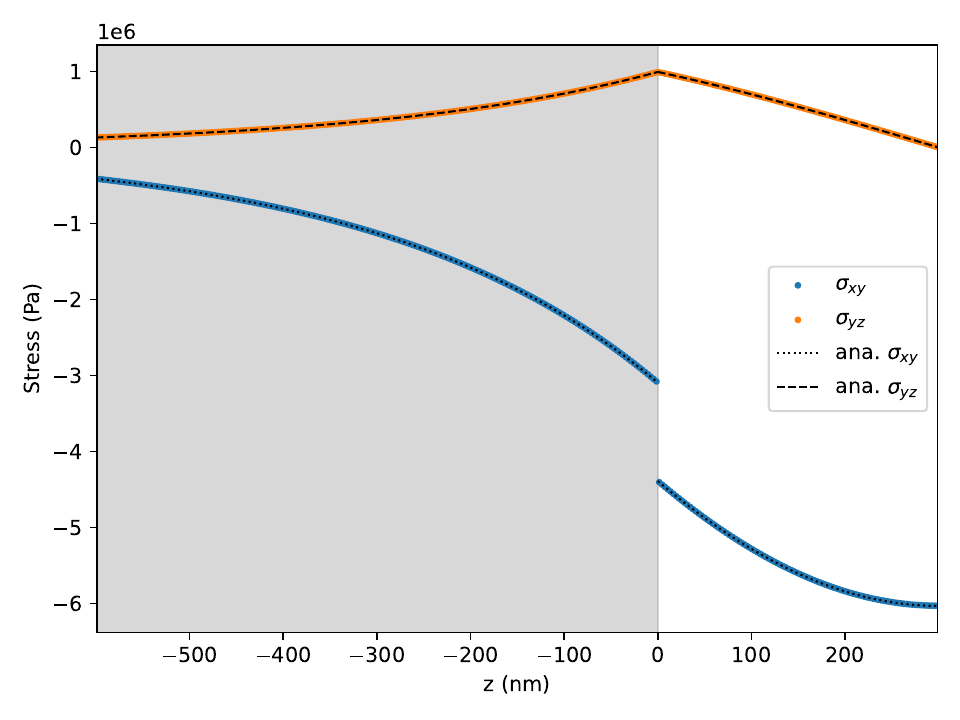}
\caption{Stress components in a SH-SAW with $\lambda=600\text{ nm}$ and $f\approx 5.26\text{ GHz}$ for a GaAs substrate with $300\text{ nm}$ of Ni on top.}
\label{fig:SH_sigma}
\end{figure}

\begin{figure}[H]
\includegraphics[width=0.7\textwidth]{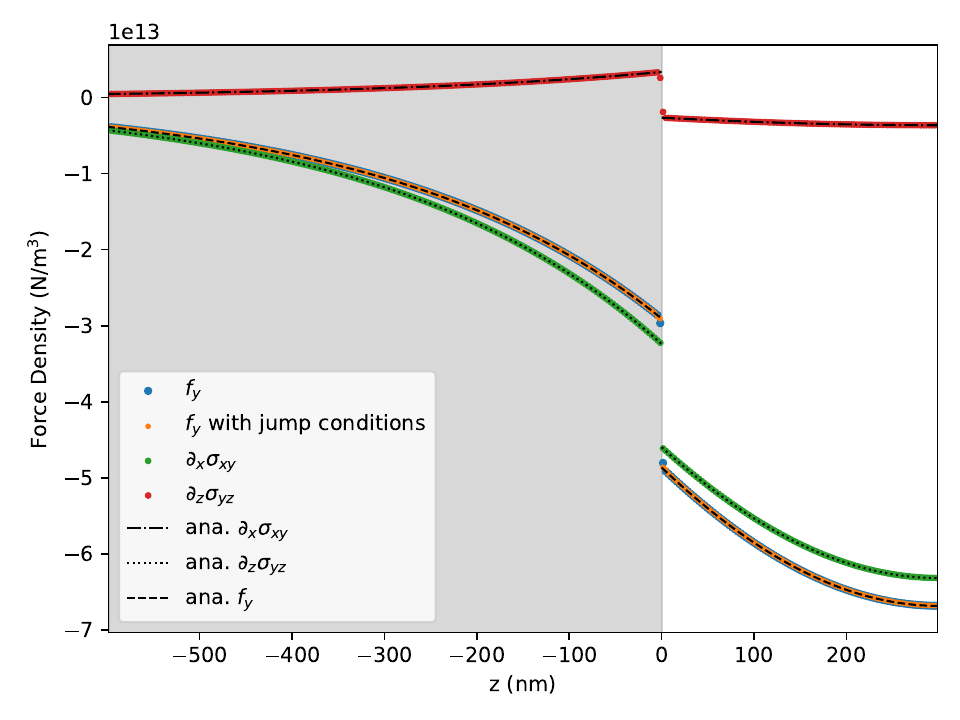}
\caption{Force density of the SH-SAW, and the constituting derivatives of $\sigma$. Using (\ref{eq:second_derivative_with_jump_conditions}) does not correctly reproduce the jump condition for continuous $\ddot{\ud}$ and introduces a small error. For comparison, a corrected scheme applicable for the SH-SAW only, according to (\ref{eq:jump_condition_SH}) is plotted.}
\label{fig:SH_f}
\end{figure}

It should be noted that even in the presence of magnetic strain, (\ref{eq:SH_compatibility}) stays approximately true as long as the magnetization is directed largely along one of the main axes. Indeed, we limit our investigation to the expected maximum of the non-reciprocity between $\phi=90^\circ$ and $\phi=-90^\circ$ \cite{Kuess_2021_PhysRevAppl}.\\
From (\ref{eq:SH_compatibility}), for a FM layer of height $h=300\text{ nm}$, we find that there is a horizontal shear mode at $f\approx 5.26\text{ GHz}$ with $\lambda=600\text{ nm}$.\\
The non-vanishing strain components and the resulting force density of this mode are shown in figures~\ref{fig:SH_sigma} and \ref{fig:SH_f}, respectively. While the jump of the strain components at the material interface is correctly reproduced, a slight deviation from the expected force density jump can be seen at the interface. This error appears because contrary to $\ud$, which we assumed to be a continuous function in our derivation, we did not require $\ddot{\ud}$ to be continuous. For the SH mode, the only non-vanishing force density component is the $f_y$ component which is composed of the derivatives $\partial_x\sigma_{xy}$ and $\partial_z\sigma_{yz}$. If we require $\ddot{\ud}$ to be continuous, we end up with a jump condition in the familiar form of (\ref{eq:jump_condition_general_form}):

\begin{equation}
\frac{1}{\rho}\left(\frac{\partial\sigma_{yz}}{\partial z}+\frac{\partial\sigma_{xy}}{\partial x}\right)\Bigg\vert_{z=z^-} = \frac{1}{\rho}\left(\frac{\partial\sigma_{yz}}{\partial z}+\frac{\partial\sigma_{xy}}{\partial x}\right)\Bigg\vert_{z=z^+}
\label{eq:jump_condition_SH}
\end{equation}

Indeed, we can see in figure~\ref{fig:SH_f}, that the jump is better approximated by applying a first derivative scheme with jump conditions by rewriting (\ref{eq:first_derivative_with_jump_conditions}) to fit (\ref{eq:jump_condition_SH}). However, this should not be considered a proper solution to this problem, since a repeated application of a second order first derivative scheme does not necessarily yield a second order approximation of the second derivative and comes with the inconvenience of effectively being a 5-point stencil scheme of low order. Instead, to properly handle this new set of jump conditions in their general form, one would need to alter (\ref{eq:second_derivative_with_jump_conditions}) by again decomposing the strains in (\ref{eq:force_density_components}) into their derivatives of $\ud$ and identifying the mixed and second derivatives of $\ud$ that take the role of the offset $B$ parameter in the jump conditions, before applying an iterative scheme similar to what we did to build up table~\ref{tab:jump_conditions_lookup}. While this can be understood well for (\ref{eq:jump_condition_SH}) in the SH mode example, the resulting workflow of this proposed scheme for the general case is quite cumbersome. For this reason, and since the error in $f_y$ appears to be small, we will instead stick to the second derivative scheme derived earlier in equation (\ref{eq:second_derivative_with_jump_conditions}).

\subsubsection{SH-SAW and SW interaction}
\begin{figure}[H]
\includegraphics[width=0.7\textwidth]{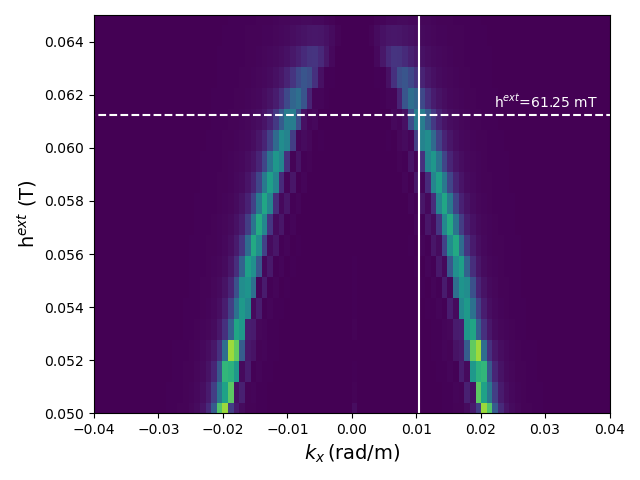}
\caption{Sweep of the external field $\bm{H}^\text{ext}=h^\text{ext}\hat{y}$ to determine the intersection of the SW dispersion with the SH-SAW mode at $600\text{ nm}$. The expected resonance field is found to be $h^\text{ext}=61.25\text{ mT}$.}
\label{fig:Ni_thick_dispersion}
\end{figure}

\begin{figure}[H]
\includegraphics[width=0.7\textwidth]{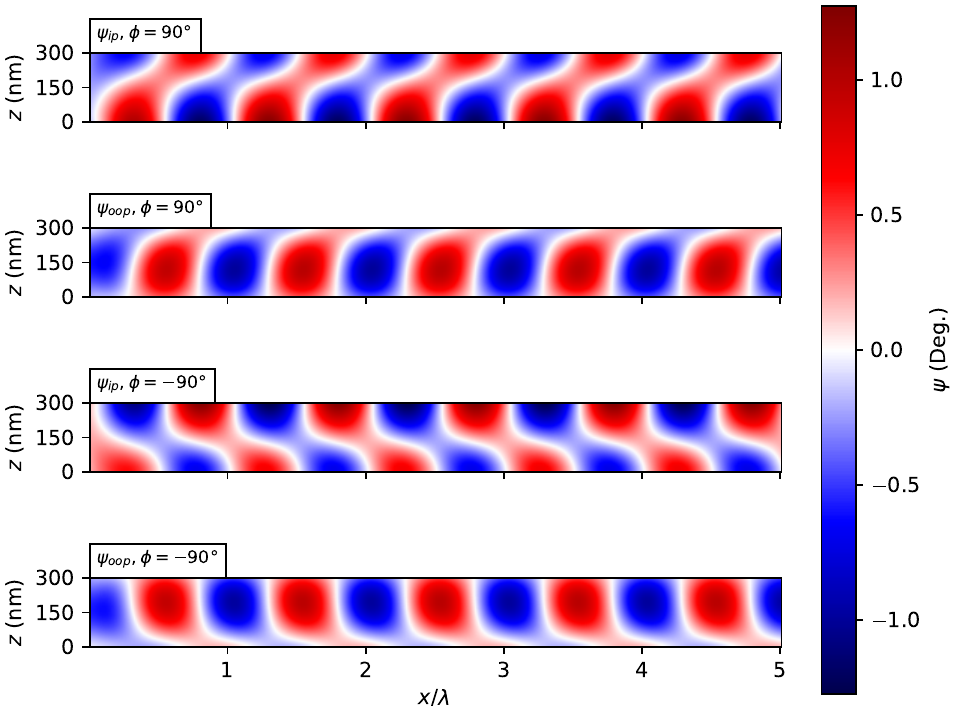}
\caption{SW excited by a SH-SAW with $f=5.26\text{ GHz}$ at $h^\text{ext}=61.25\text{ mT}$ in the Ni/GaAs system. The in-plane and out-of-plane angles of the magnetization from the initial orientation are shown for the first few wavelengths along the propagation direction ($x$-direction).}
\label{fig:SH_m_map}
\end{figure}

\begin{figure}[H]
\includegraphics[width=0.7\textwidth]{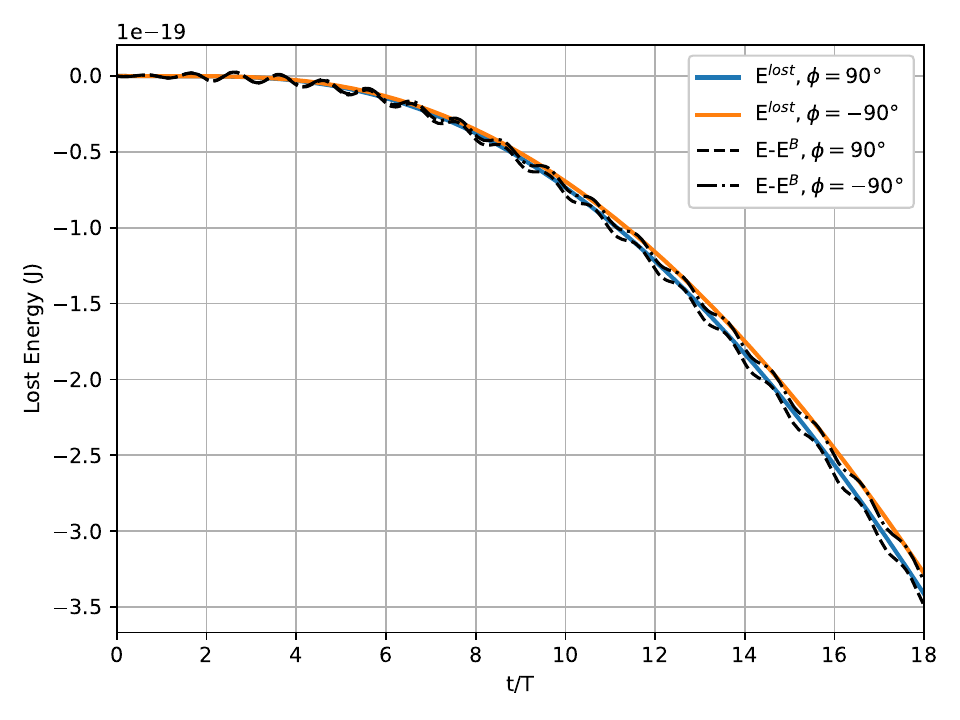}
\caption{Lost energy over time for the GaAs/Ni system, computed from the Gilbert losses and by comparing the total free energy of the in-resonance simulation to a background simulation at $\phi=45^\circ$ and $h^\text{ext}=200\text{ mT}$. The oscillating nature of the latter is attributed to a change in the SH-SAW wavelength induced by the non-zero $\varepsilon^m_{xy}$ in the background simulation.}
\label{fig:SH_loss_vs_t}
\end{figure}

\begin{figure}[H]
\includegraphics[width=0.7\textwidth]{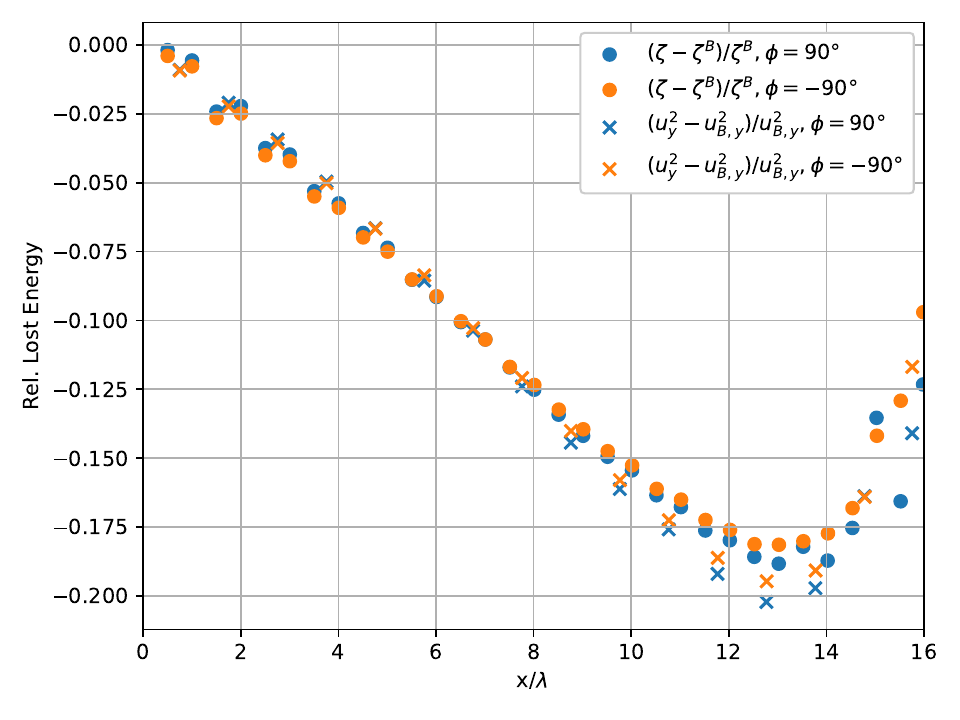}
\caption{Estimation of the SH-SAW attenuation due to magnetoelastic coupling in a $300\text{ nm}$ Ni-film, obtained from the the energy density integrated over the film thickness, and from the $u_y$ component sampled at half height of the Ni-film. The first few amplitudes found close to the right of the simulation box appear less damped since the magnetic system has not yet been fully excited.}
\label{fig:SH_loss_vs_x}
\end{figure}

Again using the sinc-pulse method \cite{Venkat_2013_IEEETransMagn} and accounting only for the dipolar field, the exchange field and the external bias field, we find an intersection of the spin-wave dispersion with our target SH-SAW at an external field of about $h^\text{ext}=61.25\text{ mT}$, as seen in figure~\ref{fig:Ni_thick_dispersion}. While the $\sigma_{yz}$ vanishes at the free surface, both $\sigma_{yz}$ and $\sigma_{xy}$ are non-zero at the GaAs/Ni interface, with a phase shift of $\pi/2$ between them. This introduces a precessional sense and thus leads to a non-reciprocity in the lost energy between the $\phi=90^\circ$ (i.e. $\bm{H}^\text{ext}=h^\text{ext}\hat{y}$) and $\phi=-90^\circ$ (i.e. $\bm{H}^\text{ext}=-h^\text{ext}\hat{y}$) case. The pattern of the magnetization's dynamic components is shown in figure~\ref{fig:SH_m_map} for the first five wavelengths. The amplitude differences of the in-plane and out-of-plane angles between the two external field directions are below $0.1^\circ$. The pattern of the spin wave indicates that the first perpendicular standing spin wave \cite{Dieterle_2019_PhysRevLett} was excited by the SH-SAW, and the phase shift between the top and bottom surface waves is due to the stray field. The dispersion calculations carried out to determine the resonant field in figure~\ref{fig:Ni_thick_dispersion} did not reveal any lower-energy modes. Under field reversal, the pattern of the spin wave in this Damon-Eshbach geometry is flipped from the top surface to the bottom surface. \\
In figure~\ref{fig:SH_loss_vs_t}, the total energy lost by Gilbert damping over time is compared to the free energy difference to a background measurement at ${(\phi=45^\circ, h^\text{ext}=200\text{ mT})}$, revealing the expected non-reciprocity between the $\phi=90^\circ$ and and the $\phi=-90^\circ$ setup. We attribute the deviation of the Gilbert losses from the oscillating difference between the background measurement and the in-resonance simulations to change in the wavelength of about $0.85$\%, that is caused by magnetic $xy$-strain in the background simulation, and a slight error introduced by the mismatching force densities at the NM/FM interface, as we discussed earlier in figure~\ref{fig:SH_f}. Overall, the two methods show good agreement.
The change in the potential energy amplitude versus the propagation distance is depicted in figure~\ref{fig:SH_loss_vs_x}. The peak values where obtained after integrating the energy density along the out-of-plane direction within the Ni-film, and the losses can be estimated well from the $u_y$ component at mid height ($z=150\text{ nm}$) of the Ni-film.  

\section{Conclusion}
We have developed a self-consistent solver for the coupled dynamics of the magnetization and the displacement, based on finite differences as an extension of the micromagnetic simulation library magnum.np. Special care was taken regarding the jump conditions at material interfaces, that arise when assuming that material parameters are constant within an FD cell. These jump conditions depend on gradient components of the displacement that run parallel to the interface, making it necessary to calculate the mechanical strain and elastic force density iteratively.\\
We then discussed a selection of computational experiments, that illustrate the implications of these jump conditions in the context of spin-wave and acoustic-wave coupling: First, we have shown how the boundary terms at an FM/NM interface results in mechanical strain within the FM material. Next, we have investigated how a shear bulk mode is attenuated in a ferromagnet, after injecting it using Neumann boundary conditions. We observed that no damping occurs if $\m$ is perpendicular to the propagation and the shear direction. For an initial magnetization along the shear direction, we calculated the injected energy from the boundary conditions and showed that all energy lost from the system is due to the Gilbert damping.\\
Finally, we investigated the attenuation of SAWs in two systems. First, for a Rayleigh mode in a magnetic thin-film structure, as experimentally measured in~reference~\cite{Kuess_2021_PhysRevAppl}, using fitted parameters to approximate the wave profile in LiTaO$_3$ near the surface. We found the expected asymmetry in the transmission losses with respect to the external field angle. Then, we carried out a similar simulation for a horizontal shear mode in a magnetic thick-film structure, where the height of the nickel film was half the wavelength of the horizontal shear mode. At this thickness, the magnetization can no longer be assumed homogeneous in the out-of-plane direction. Again, the expected asymmetric behavior in the transmission losses was found.\\
These results not only show the versatility and robustness of the presented self-consistent magnetoelastic solver, but hopefully also provide a set of representative validation problems to the community.

\section*{Acknowledgement}
This research was enabled by the financial support of the Austrian Science Fund (FWF) I6068 and the Deutsche Forschungsgemeinschaft (DFG, German Research Foundation) 504150161.

\bibliographystyle{bib/thesis.bst}
\bibliography{bib/lit.bib}
\end{document}